\documentclass[preprint]{aastex}

\received{}
\revised{}
\accepted{}

\begin{document}

\title{A Study of 3CR Radio Galaxies from z = 0.15 to 0.65.  I.  Evidence
for an Evolutionary Relationship Between Quasars and Radio Galaxies}

\author{Michael Harvanek\altaffilmark{1,2}, E. Ellingson\altaffilmark{2},
\& John T. Stocke\altaffilmark{2}}
\affil{Center for Astrophysics and Space Astronomy, CB 389
University of Colorado, Boulder, Colorado, 80309-0389, \\
electronic mail: harvanek, stocke, e.elling@casa.colorado.edu}

\and

\author{George Rhee\altaffilmark{2}}
\affil{Physics Department, University of Nevada, Las Vegas, NV, 89195 \\
electronic mail: grhee@dawg.physics.unlv.edu}

\altaffiltext{1}{Current Address: Apache Point Observatory, 
2001 Apache Point Rd.,
 PO Box 59, Sunspot, NM, 88349-0059, email: harvanek@apo.nmsu.edu}
\altaffiltext{2}{Visiting Astronomer, Kitt Peak National Observatory,
National Optical Astronomy Observatories, which is operated by the
Association of Universities for Research in Astronomy, Inc. (AURA)
under cooperative agreement with the National Science Foundation.}


\begin{abstract}

Deep optical images have been gathered for a nearly complete sample
of radio galaxies from the Revised 3rd Cambridge (3CR) Catalog
in the redshift range $0.15 < z < 0.65$.  Total and nuclear
magnitudes and colors have been extracted.  The richness of the
galaxy clustering environment has also been quantified by calculating the
amplitude of the galaxy-galaxy spatial covariance function
($B_{gg}$), 
showing overdensities ranging up to Abell class 0-1 clusters.
These optical data are compared to similar data from an
existing sample of radio-loud quasars in the same redshift range for the purpose
of determining the relationship between radio galaxies and quasars.
In the range $0.15 < z < 0.4$, we find that quasars and radio galaxies
have significantly different environments in that only radio
galaxies are found in rich cluster environments.
This comparison appears to rule out the hypothesis that all quasars
are radio galaxies viewed from a particular angle at the $97\%$
confidence level ($99.6\%$ confidence level if N-galaxies are
considered quasars). The existence of quasars in clusters at
$z > 0.4$  supports the hypothesis that some radio-loud quasars
may dim with time and evolve into radio galaxies
with an e-folding time of $\sim 0.9$ Gyr.
A compatible
scenario is presented for this evolution in which the
quasar dims due to the absence of low velocity interactions
between the quasar host and companion galaxies which trigger
quasar activity and/or a diminishing fuel supply caused by the
more effective gas ``sweeping'' of a growing intracluster medium.

\end{abstract}


\section{Introduction}
Radio galaxies are one of the many
types of active galactic nuclei (AGN) recognized today.  Currently,
the list of AGN types includes quasars, QSOs, radio galaxies (broad
and narrow lined), Seyfert galaxies (types I and II) and
blazars.  The observational properties of the different types
of AGN vary considerably, and although all are thought to be
driven by a supermassive ($> 10^6$ solar mass) black hole at the
nucleus of their host galaxy, the relationships among the
various types of AGN are not well understood.
The currently accepted
model for radio-loud AGN invokes a supermassive black hole that powers
a relativistic jet, creating Doppler-boosted, beamed radiation.
Differences in
the amount of obscuration of the AGN, orientation of the jet
with respect to the line of sight, motion of the jet, properties
of the black hole and evolution have all been used in various
combinations 
to explain the differences in the
properties of the various AGN types.
However, to date, no one model has been able to explain them all.
In fact, the ``unification''
of this diverse group of objects is currently one of the primary
goals of extragalactic astronomy.

In an effort to 
further explore the unification of AGN, this work will
examine one of the more controversial AGN relationships: that between
quasars and radio galaxies. 
The radio sources associated with quasars
are generally quite similar in both luminosity and morphology to the
radio sources associated with the more powerful radio galaxies;
typically, both are Fanaroff-Riley Type 2, or FR2, sources (\citealt{FR}).
Some differences in the radio properties do exist however.  Radio
galaxies have, on average, weaker radio cores (\citealt{Miley},
\citealt{Owen}), exhibit weaker radio jets (\citealt{Owen}), and can have
larger linear dimensions (\citealt{Miley}) than quasars.
The spectra of quasars are characterized by broad emission lines.
The spectra of a few
radio galaxies show broad emission lines while most show only narrow emission
lines."
Quasars are luminous in
X-rays (e.g., \citealt{Zam}; \citealt{Worrall}) whereas radio galaxy AGN
are comparatively dim (e.g., \citealt{FeigBerg}).  While many are found in
elliptical host galaxies, a significant number of quasars (\citealt{Smith};
\citealt{Hutch}; \citealt{StockMac}) and powerful radio galaxies (\citealt{Heck};
\citealt{Hutch}; \citealt{deKoff}) are found in hosts with peculiar
optical morphologies, indicating interacting and/or merging systems.
Finally, more
luminous quasars and radio galaxies (e.g., those from the 3CR catalogs of
\citet{LRL} and \citet{Spinrad}) have remarkably
similar redshift distributions at $z \gtrsim 0.4$ (\citealt{Longair};
\citealt{Barthel}).
All these various relationships between quasars and radio galaxies,
along with their differing projected physical sizes and optical 
polarization properties (see \citet{Barthel} and
references therein) and evidence for the preferred orientation of some quasars
led \citet{Barthel} to propose that quasars and FR2 type radio
galaxies are the same type of object viewed from a different orientation
angle.  Since their properties suggest that quasars are beamed
toward us, this hypothesis
proposes that quasars are FR2 type radio galaxies with their jets
oriented closer to our line of sight (within $\sim 45 ^\circ$).

However, the work of Yee \& Green (1987, hereafter YG),
Ellingson, Yee \& Green (1991a, hereafter EYG) and Yee \& Ellingson 
(1993, hereafter YE93) 
suggests
another possibility.  These studies indicate that while up to
one third of optically bright, radio-loud quasars at $z \sim 0.6$ are
found in Abell class 0 or richer clusters,
a much smaller fraction of lower redshift quasars are found in rich environments.
In addition, they found that the brightest quasars which inhabit clusters at $z \sim 0.4$
are several magnitudes fainter than quasars in clusters at higher redshifts.
The only radio-powerful AGN found in clusters at low redshifts are radio galaxies.
In contrast, the fraction and luminosity of
the brightest quasars in poorer environments change relatively little over
these epochs.
This suggests that environment  strongly affects
quasar evolution, and that quasars are rapidly disappearing from
rich clusters over this redshift range. 
Based on their findings, EYG proposed that quasars may dim
with time and eventually fade to become radio galaxies.  This will be
referred to as the evolutionary hypothesis of EYG. 
Note that this hypothesis is not wholly inconsistent with the beaming
phenomenon-- only with viewing angle being the {\it sole} explanation
for the different AGN classes.  In the evolutionary hypothesis,
the nuclear properties of some individual objects may be affected by beaming, but 
there is also a fundamental evolutionary connection between luminous quasars
and radio galaxies.

One key to testing these competing hypotheses is the 
galaxy environments of the FR2 radio galaxies.
Assuming that the richness of the
galaxy environment is independent of the orientation angle to the
line of sight (i.e., the cluster richness remains the same
regardless of the angle from which it is viewed), if quasars and
FR2 radio galaxies are the same type of object viewed from a different
angle, the galaxy environments of quasars and FR2 radio galaxies at
the same redshift should be similar. 
EYG found essentially no optically bright quasars in clusters at $0.15 < z < 0.4$;
thus, the orientation
angle hypothesis of \citet{Barthel} predicts that no FR2 radio
galaxies in this redshift range should be found in clusters.
However, if quasars evolve into radio galaxies, the galaxy
environments of radio galaxies at $0.15 < z < 0.4$ should
be similar to that of quasars at some earlier epoch.  Since
EYG found quasars in both rich and poor environments at $z > 0.4$,
their evolutionary hypothesis predicts that the
percentage of FR2 radio galaxies found in rich environments
at $0.15 < z < 0.4$ should be similar to the percentage of quasars
found in rich environments at some epoch $z > 0.4$.

However, consistent data on the galaxy environments of FR2
radio galaxies at $0.15 < z < 0.4$ has been lacking. 
Although a number of studies of the environments of radio galaxies have
been conducted previously (\citealt{Lilly}; \citealt{Prestage}; Yates,
Miller \& Peacock 1989; hereafter YMP, Hill \& Lilly 1991;
hereafter HL; \citealt{AEZO93}; \citealt{Z97}), much of the data
collected were for radio galaxies at higher or lower redshifts.
(Note that data for radio galaxies at
$z < 0.15$ are abundant.  However, since quasars are not found
at $z < 0.15$, this data cannot be used to test Barthel's
hypothesis.)  The studies that did provide adequate coverage
over the relevant redshift range (\citealt{AEZO93};
\citealt{Z97})
used different filters and quantified the
richness of the galaxy environment in a somewhat different
manner.

Thus, we obtained deep optical images of a sample of 39 radio-powerful 
3CR AGN in this redshift range to determine the properties of
the AGN nuclear sources and their environments.
In Section 2 the radio galaxy and quasar samples are presented.
The observations, reductions, photometry and other data processing are
discussed in Section 3.  
In Section 4 the $B_{gg}$ parameter, which quantifies the richness of
the clustering environment, is presented in some detail.  Incorporation
of other clustering data is discussed and the environmental data is
tabulated.
In Section 5 the properties of the radio galaxies are compared to the
properties of quasars and the results of this comparison are used to
determine whether the orientation angle hypothesis of \citet{Barthel}
or the evolutionary hypothesis of EYG can better explain the
relationship between quasar and radio galaxy environments.  A summary
of the results is given in Section 6 and a scenario which accounts
for them is proposed.
Values of H$_0$ = 50 km s$^{-1}$ Mpc$^{-1}$ and q$_0$ = 0 (or 0.02)
are assumed throughout this work for consistency with EYG.
In a companion paper (\citealt{paper2}, Paper 2 hereafter) we
study the extended radio structure of the sources in this sample.

\section{The Samples}
\subsection{The Observed Sample}
The sample chosen for this study was drawn from the 3CR radio
galaxies and quasars with $0.15 < z < 0.65$ and $|b| \ge 15^\circ$
that are listed in the Revised 3C Catalog of Radio Sources of
\citet{SSS} as updated by \citet{Spin85} and \citet{Spinrad}.
This catalog
was chosen as the source of our sample for several reasons.  Having
originated from the 3CR radio catalog (a list of more than 300 of
the brightest radio
sources observed at 178 MHz by \citet{Bennett}), this catalog ensures
that any source at $z > 0.15$ chosen from it will be comparable in
radio power to the sources in the quasar comparison sample of
YE93 (see below).
Furthermore, since the catalog is nearly complete (for sources having
178 MHz flux $\ge 10.9$ Jy on the scale of \citet{Baars}) and
almost entirely identified ($>$ 91\%; \citet{Spinrad}, and
objects with $|b| \ge 15^\circ$ are essentially 100\% complete)
it provides a very comprehensive listing of such sources.
Choosing sources from only one flux-limited catalog also helps
ensure a uniform sample.  
The galactic latitude was restricted because the
large number of stars and the possibility of uneven extinction near 
the galactic plane make the measurement
of the galaxy environment more uncertain.

Table 1 shows the sample taken from the 3CR catalog and tabulates
some of the more useful properties.
There are 66 radio galaxies and 14 quasars listed.  However,
one of the galaxies, 3C\,258, was discovered to contain what
is believed to be a distant background quasar in its spectrum
(A. Dey, private communication).  Since the angular size
of 3C\,258 is quite small for a source at a redshift of $z = 0.165$,
it is likely that the radio source is associated with the
background object.  Thus, this object has been removed from
the sample and is excluded from any of the analysis presented
in this study.  

Listed in Table 1 along with the sources are their
optical positions, redshifts, 178 MHz flux densities on the scale of
\citet{Baars}, and spectral indices between 178 and 750 MHz.  (The spectral
index $\alpha$ is defined here in the sense $S \propto \nu^{-\alpha}$.)
The optical positions and redshifts were taken from \citet{Spinrad} except
for the optical positions of 3C\,225B and 3C\,435A, which were taken from
\citet{4m} (from the cross on the radio map) and \citet{r200},
respectively.  The 178 MHz flux densities were taken from
\citet{LRL} if available.
If not available in \citet{LRL}, the values given in 
\citet{KPW} were
used for most other sources.  The 178 MHz flux densities for
3C\,99, 3C\,258, and 3C\,306.1 were taken from
\citet{Gower} and that
for 3C\,268.2 was taken from \citet{Pilk} (all available from the NASA/IPAC
Extragalactic Database, 1997)
because their values in \citet{KPW} are likely contaminated by nearby
sources.  The guidelines for choosing the most appropriate 178 MHz flux
density for any given source are discussed in detail in \citet{LRL}.
All fluxes were adjusted to the scale of \citet{Baars} using the
corrections given in \citet{LaiPea}.
Spectral indices were computed using the
178 MHz flux densities given here and 750 MHz flux densities from
\citet{KPW}.
Comparison of our spectral index values to
those of \citet{LRL} shows perfect agreement for all 48 objects common
to both samples.  For one source, 3C\,435A, the 178 MHz flux density and
the spectral index were not directly available because the separation of
3C\,435 into the two unrelated sources, 3C\,435A and 3C\,435B, occurred
after the work of \citet{KPW} and we were unable to find these values
in more current literature.  Since no other data were available, the
spectral index of 3C\,435A listed in Table 1 is that of the
combined source, 3C\,435.
The 178 MHz flux density was computed using the 1500 MHz flux density
of 3C\,435A from \citet{r200} and a spectral index for the
combined source calculated from fluxes at the appropriate frequencies
taken from \citet{KPW}.

Also included in Table 1 are the rest frame luminosity (i.e., power) at
178 MHz, the log of this rest frame luminosity, the optical spectral
type of the source (i.e., the ``optical class'') and the object type of
the source.  The rest frame luminosity was calculated from the 178 MHz
flux density, the spectral index and the redshift, assuming a Friedmann
cosmology with $H_0$ = 50 km s$^{-1}$ Mpc$^{-1}$ and $q_0$ = 0.  All
sources in this sample have a rest frame luminosity above (most well
above) the nominal boundary between FR1 and FR2 type radio sources
(\citealt{FR}).  The radio morphology of these sources mostly confirms this FR2
luminosity classification (see also Paper 2 for a more detailed analysis and
discussion). The same is true of the quasar comparison
sample of YE93 discussed below.
We classify each object as ``broad-line" (B), ``narrow-line'' (N) or
``low-excitation" (E).
Except for 3C\,234 and
3C\,381, the optical type was taken directly from \citet{Jack} who
use a  slightly different notation than that used here.  Their ``quasar/weak
quasar'' (Q/WQ) classification is equivalent to our broad-line (B) type.
Similarly, their ``high-excitation galaxy'' (HEG) corresponds to our
narrow line (N) type and their ``low-excitation galaxy'' (LEG) is the
low-excitation (E) type.  The references for the optical spectra on which 
these classifications are based can be found in \citet{Jack}.  The optical
type for 3C\,234 and 3C\,381 was taken from \citet{4l} who also provide
the references for the optical spectra from which these classifications
are determined.
The object types are galaxy (GAL), N-galaxy (N) and quasar (QSO)
and were taken directly from \citet{Spinrad}.

\subsection{The Quasar Comparison Sample}
For the comparison study, we have utilized the quasar sample of YE93
which contains 65 quasars with $z < 0.65$ and $|b_{II}| \ge 30^\circ$, of which
10 are 3CR sources and most of the remainder are from the 4C and Parkes catalogs. 
Values of redshift, absolute nuclear B magnitude and $B_{gq}$ (quasar-galaxy
two point correlation function amplitudes; see Section \ref{bgg}) are provided
by YE93 and for the comparison study we adopt their values directly.
When combined with the 3CR sample discussed above, there are a total of
69 quasars and 65 radio galaxies in this study.

\section{Observations and Reductions}
All optical imaging data were obtained at the 0.9 m and 2.1 m
telescopes of the National Optical Astronomy Observatories
(NOAO) at the Kitt Peak
National Observatory (KPNO).  The lower z subsample ($0.15 < z < 0.35$) was observed
with the smaller telescope 
(FOV $\sim 23'$ on a side) while the higher z subsample 
($0.35 < z < 0.65$) was
observed with the larger telescope (FOV $\sim 5'$ on
a side).

The optical observations are listed in Table 2.  The
Gunn $r$ and Gunn $g$ filters were used for the observations because
they match those of the background galaxy counts (\citealt{YGS86},
as updated by H. Yee, private communication) and allow
for a direct comparison with the quasar work.
The integration times are typically 1800 sec in $r$ and
3600 sec  in $g$ to yield a completeness magnitude 2-3
mags dimmer than $M^*$ ($M_r^* \sim -22.0$; the combination
of k-correction plus moderate luminosity evolution keeps this value
approximately constant over the redshift range of interest.)
The $5\sigma$ limiting magnitude for each field is
listed in Table 2.  The completeness magnitude
is estimated to be $\sim 0.8$ mag brighter than the $5\sigma$
limiting magnitude (\citealt{Y96}).
 
The identification and classification of all objects in each field and
the calculation of an instrumental magnitude for each object were
performed using the Picture Processing Package (PPP) software developed
by \citet{Yee91}, and further described in \citet{Y96}.
To calibrate the photometry, standard stars from
\citet{Kent} were observed at several different times
during each night of each of the observing runs.
Observational uncertainties are typically 0.03-0.1 mag.
The observed total apparent magnitudes were corrected
for Galactic reddening by using $(A_\lambda/A_V)$
given by \citet{CCM} with the currently accepted value of
$R_V = 3.1 \pm 0.1$ and values of $E(B-V)$ for each field
taken from \citet{BurstHei}.
Gunn $r$  absolute  magnitudes were k-corrected using the
values given in \citet{Sebok} for E/S0 type galaxies as a function of $z$
and Gunn $g$ magnitudes were corrected using the tables from \citet{Fuku}.
Total magnitudes (apparent and absolute) and
observed $g-r$ colors for the radio galaxies
are given in Table 3.  Those for the
quasars are found in Table 4.

Total magnitudes marked with a colon
were not obtained from our data.  They are estimates based on 
the brighter of $m$ and
$m_{15}$ (total magnitude and magnitude inside 15 kpc, respectively) given in \citet{deKoff}.  These magnitudes
were obtained with the Hubble Space Telescope (HST) using the 
broad-band F702W filter  
and were converted to Gunn $r$ 
using the conversions given in
\citet{Fuku} for E type galaxies as a function of $z$.
Due to low exposure times and the combination
of the extremely high resolution and small pixel size of HST, the
magnitudes taken from \citet{deKoff} may be dimmer than the actual
total magnitudes and so entries marked with a colon in Table 3
should be used with caution.

To estimate the nuclear magnitudes for the radio galaxies,
we used $m_1$ 
(the magnitude within a 1 kpc radius for H$_0$ = 75 km s$^{-1}$ Mpc$^{-1}$,
q$_0$ = 0.5, which corresponds to 0.2-0.36" for this redshift range) from \citet{deKoff},
appropriately corrected to obtain $M_B$ for H$_0$=50 km s$^{-1}$ Mpc$^{-1}$, 
assuming a typical power-law AGN spectrum 
with $\alpha = -0.5$ (YE93), equivalent to a $g-r$ color of -0.11 mag.
However, these $m_1$ magnitudes
will contain a small contribution from the host galaxy, although the
exact amount is unknown without a detailed knowledge of the inner
surface brightness profile. If a deVaucouleurs
profile (\citealt{deVauc}) is assumed, the galaxy contribution will be typically a few tenths of
a mag at most. 
Therefore, we have used $m_1$ mags as an estimate of the
nuclear magnitudes 
for all those galaxies in Table 1 whose optical class is N or
B. For N or B class radio galaxies, ground-based spectra through apertures considerably 
larger than 1 kpc are dominated by non-thermal continuum. 
Additionally, a visual
inspection   of the spectra 
in \citet{Tadhunter} for those galaxies
in Table 1 left unclassified by \citet{Jack}
finds their ground-based spectra likewise non-thermally dominated.
The spectra of galaxies classified as E type, in some cases (e.g., 3C\,348), 
but not all 
show optical
spectra dominated by host galaxy starlight (\citealt{Tadhunter}).
To be conservative, we have
listed $m_1$ mags  as a limit on the nuclear brightness of galaxies
of type  E in Table 3. This relative inaccuracy in our nuclear magnitude
estimates does not affect our final results. 

\section{Clustering Analysis}\label{bgg}
This work utilizes $B_{gg}$, the amplitude of the galaxy-galaxy spatial
covariance function, as a quantification of the richness of the clustering
environment around a given object.  Given a cosmology, an assumed evolution
of the galaxy luminosity function and measured mean background
galaxy counts, this parameter reflects the galaxy overdensity
around a given object, correcting for the expected spatial and
luminosity distributions of field galaxies and of the associated cluster
galaxies at the redshift of the object.

This parameter is described in
detail in \citet{LS79} and the specific technique used in obtaining the
actual values of $B_{gg}$ is described in \citet{YG87} and EYG.
Briefly, the technique is as follows.
All galaxies within 0.5 Mpc of the radio source
brighter than some magnitude are counted.  Then the
expected number of background galaxies in that area down to the same
magnitude is subtracted.  The number of excess galaxies is normalized
to an evolved galaxy luminosity function at the redshift of the object
and then converted to $B_{gg}$ assuming a form for the spatial
distribution of galaxies.

As explained in EYG, the 0.5 Mpc counting radius is chosen
to provide good contrast between cluster and foreground/background galaxies and
to minimize the effects of the variation of the actual spatial
distribution of the cluster galaxies from the assumed form.
The magnitude limit for galaxy counting is determined for each field
individually and is taken to be the brighter of the ``completeness''
magnitude, $m_{comp}$, or M$^*_r$ + 2.5.
Background galaxy counts
are those of Yee et al. (1996,1998).
The evolving galaxy luminosity function used to normalize the
excess galaxies includes 
moderate galaxy evolution of M$_*$($z$) $\sim z$ (EYG).
Finally, the distribution of the excess
galaxies is assumed to be the standard cluster galaxy power law
($r^{-\gamma}$ where $\gamma = 1.77$).  This originates from the
angular distribution of \citet{SP78} and was used by both \citet{YG87}
and EYG.  \citet{YG87} show this assumption is consistent
with their results.  

It should be noted that although $B_{gg}$ is a measure of the excess
galaxies around a given object, a large value of $B_{gg}$ does not
prove absolutely that there is a cluster around any individual object.
The excess galaxies could be due to
an anomalous overdensity of foreground and/or background galaxies in
that region of the sky.  Although galaxy colors can be used as an
indication of the redshift of the excess galaxies, only spectroscopic
observations can completely confirm cluster membership.  For those
who are more familiar with the cluster richness classification of
\citet{A58}: $B_{gg}$ values of 600 $\pm$ 200, 1000 $\pm$ 200,
1400 $\pm$ 200, and 1800 $\pm$ 200 Mpc$^{1.77}$ are comparable to
Abell richness classes 0, 1, 2, and 3, respectively (\citealt{YL-C}).
EYG chose $B_{gg} = 500$ Mpc$^{1.77}$ to define the boundary
between rich and poor environments because, at the time, it was thought
to be the division between Abell richness classes 0 and 1
(\citealt{Prestage}, 1989). We adopt this boundary for consistency with
the earlier work and show below that this choice does not affect
our results significantly.

The total uncertainty in $B_{gg}$ is given by
$\Delta$$B_{gg}/B_{gg} = (N_{net} + 1.3^2N_{b})^{1/2}/N_{net}$
(\citealt{YL-C}) where $N_{b}$ is the expected number of background galaxy
counts and $N_{net} = N_{total} - N_{b}$ is the number of excess galaxy
counts.  This uncertainty is obtained by adding two terms in quadrature.
The first term, $(N_{net})^{1/2}/N_{net}$, is the internal statistical
error in sampling the associated cluster galaxy luminosity function
using a finite number of galaxies.  The second term,
$1.3(N_{b})^{1/2}/N_{net}$ (EYG), is the error due to the
uncertainty in the background galaxy counts.  The factor of 1.3 is
an empirical value that accounts for the clustering of the background
galaxies which causes the error in the background galaxy counts to
deviate from a Poisson error.  This factor is discussed in detail in
\citet{YGS86}.  The error in $B_{gg}$ is often quite a substantial
fraction of the $B_{gg}$ value due to the small number of ``excess'' galaxies
relative to the expected number of field galaxies.
Two of the quasar fields we observed (3C 323.1 and 3C 351) were also
observed by YE93.  A comparison of our $B_{gg}$ values with
theirs shows differences within the uncertainties that are
easily attributable to differences in
the limiting magnitudes.

\subsection{Incorporation of Other Clustering Data}
We have measured $B_{gg}$ for 39 of the 79 fields in our 
sample.
However, measures of the excess galaxies surrounding many additional
objects exist in the literature in one form or another
and have been incorporated into this study.
Table 5 provides a comparison between our method of measuring $B_{gg}$
and the other methods of measuring the number of excess galaxies.
The differences between methods and the conversions to our
$B_{gg}$ values are discussed below for each method listed in
Table 5.

\subsubsection{Yee \& Ellingson (1993)}
Nuclear magnitudes and B$_{gg}$ values for 63 quasars (8 of which
are 3CR sources) with $0.15 < z < 0.7$ were taken directly from
YE93, which includes the samples used in EYG.
These $B_{gg}$ values were obtained in a manner
almost identical to our own.  Some of their values (those taken
from YG and \citet{YG84}) were calculated using q$_0$ = 0.5
with a slightly different luminosity function (that of \citet{Sebok}
rather than the luminosity function of \citet{KE85} that
we used; see YG) and with slightly different background
counts, but these differences are minor.
We have thus taken their values without any correction.
As mentioned above, comparisons of $B_{gg}$ values for objects in
common agree to within the uncertainties.

\subsubsection{Zirbel (1997)}
Zirbel (1997) observed a sample of radio galaxies
(of which 27 are in the 3CR sample in our redshift range)
and quantified the galaxy environment using
N$^{-19}_{0.5}$ (or ``richness''):
the number of excess galaxies brighter than M$_V$ = --19 within
a 0.5 Mpc radius of the object.  The main differences between this
method and our method are that the images are in V rather than
Gunn $r$, the limiting magnitude is fixed for all fields, and galaxies
that are -0.6 mag bluer or 0.2 mag redder than an elliptical galaxy
of the same absolute magnitude are excluded from the excess galaxy
counts.  
Since these galaxies are red,
faint galaxies that would be detected in Gunn $r$ may not be detected
in V.  The fixed limiting magnitude of M$_V$ = -19 means that we
tend to count galaxies down to fainter magnitudes, especially at lower
redshifts (z $<$ 0.3), but this should be accounted for in the
normalization.  The effect
of excluding galaxies with anomalous colors is to reduce the
number of excess galaxies.  So, in general, we expect the number
of excess galaxies measured by \citet{Z97} to be less than ours.
Since $B_{gg}$ is directly proportional to the number of excess
galaxies, we expect the N$^{-19}_{0.5}$ values of \citet{Z97} to be
converted to our $B_{gg}$ values by a multiplicative constant.
Since a correction for the luminosity evolution of the galaxies 
is made by \citet{Z97} (see \citet{AEZO93} for
details), this multiplicative constant is expected to be the same
for all fields regardless of redshift.  However, the differences
between the two methods discussed above will cause a slight variation
in this multiplicative constant from field to field and so there will
be some scatter around the actual conversion value.

In order to determine this conversion constant, a weighted least
squares linear fit of Zirbel's ``richness'' values to our $B_{gg}$
values was performed using 13 objects for which we both had data.
The fit was forced through the point (0,0) because zero excess galaxies
should give $B_{gg}$ = 0.  The resulting conversion is
$B_{gg}$ = 38N$^{-19}_{0.5}$.  The fit and the data points with their
error bars are shown in Figure 1.  The 1$\sigma$ error bars for all but
3 of the points overlap the fit and the farthest of these 3 points
(3C 348) lies within 2$\sigma$ of the fit.

\subsubsection{Yates, Miller \& Peacock (1989; YMP)}
This work includes data from 14 3CR radio galaxies in our redshift range.
The $B_{gg}$ values are the same quantity that
we calculate.  However, their values are not directly comparable to
ours due to differences in methodology and so a conversion is
still required.  While their model 2b is the closest to our work (q$_0$ = 0,
a \citet{KE85} luminosity function, YG normalization and evolution),
they do not distinguish between stars and galaxies
when counting objects, the area over which they count objects varies
from field to field and their computation of the completeness
magnitude is different from ours.  Although  the inclusion of stars to their
number counts
sounds like a crucial difference, it
should be accounted for by their ``local'' (5$'$ to 10 $'$ offset)
measurement of background objects (stars and galaxies).  This 
assumes that there are no large differences in the stellar number
density between an object field and its offset frame.  However,
not directly eliminating the stars from the analysis increases
the random uncertainty in $B_{gg}$.  
\citet{YMP89} also take their completeness magnitude to be at the peak of their
galaxy number distribution whereas we take it to be where the
galaxy number distribution begins to drop below the expected
linear form.  This will cause \citet{YMP89} to overestimate their
completeness magnitude, giving them lower counts and a lower
value of $B_{gg}$ than ours for the same completeness magnitude.
However, the non-uniform variation of both our completeness magnitude
and that of \citet{YMP89} from field to field prevents this from appearing 
as a systematic difference in the objects we both observed.

We can attempt to model the effects of the
differences between our method and that of \citet{YMP89}
as an additive offset between the two $B_{gg}$ values.
Thus, we look for a conversion of the form:
$\{$our $B_{gg}\}$ = $\{$YMP $B_{gg}\}$ + constant.
Note, however, that the variation of the differences in
the two methods from field to field may cause this offset to vary.
An average offset of 70 was determined from 5 of the 7 objects for
which we both had data.  The remaining two objects (3C 346 and 3C 348;
the uppermost points of Figure 2)
were not used in the determination of this offset because they both
had extremely small counting areas when compared to ours (smaller by
a factor $\sim$ 3) and so represented extreme rather than typical
differences.  A comparison of this ``average offset'' conversion
to a weighted least squares linear fit to the same 5 points is
shown in Figure 2.  The offset conversion is given by the solid line.
The two lines agree to within their error bars (y = x + 70$\pm$147 vs.
y = 0.91$\pm$0.54 $\times$ x + 105$\pm$73) and when the size of the error
bars of the points is considered, the difference between the two
fits is quite small.
Only 3C 348 is noticeably
offset from the fit and it is still well within 2$\sigma$.

\subsubsection{Hill \& Lilly (1991; HL)}
Hill \& Lilly (1991) observed a sample of radio galaxies
at $z \sim$ 0.5, 13 of which are included in this study.
Their measurement of N$_{0.5}$ is derived from the number of excess
galaxies within a 0.5 Mpc radius of the object within the magnitude
range m$_1$ to m$_1$ + 3 where m$_1$ is the magnitude of the object.
We note that this method does include
a built-in correction for galaxy evolution and k-corrections, as long
as the radio galaxies can be considered typical of cluster galaxies.
This quantity is computed using q$_0$ = 0.5.  Another listed
quantity, N$^M_{0.5}$, is the same as N$_{0.5}$ except that the mean
m$_R$-z relation for the object is used as m$_1$ rather than the
magnitude of the object itself.  A third tabulated quantity, N$^0_{0.5}$,
is the same as N$_{0.5}$ except that q$_0$ = 0 was used in the
calculation.  Although this
third quantity is a closer match to our value of q$_0$, the second
quantity is more consistent with our method of determining the
limiting magnitude.
Using the N$^M_{0.5}$ values, the differences
between our method and that of \citet{HL91} are the different value
of q$_0$, the slightly different optical waveband, variations in the
limiting magnitude from field to field, and the fact that they do not
distinguish between stars and galaxies.
The effects of varying q$_0$ are
negligible compared with the uncertainties. 
The difference in optical waveband
is small (R vs. Gunn $r$), and should not systematically affect the results
as long as both field and background counts are
observed in the same band.
Varying the limiting magnitude should have no systematic
effect on the results, but can produce statistical variations in 
different measurements for individual objects, depending on how deep
the luminosity function is sampled.
Finally, as with \citet{YMP89} above, 
the local measurement of background objects should account for stars
not being removed but the potential for much larger
errors is increased.

As with the N$^{-19}_{0.5}$ values of \citet{Z97}, we expect the conversion
of the \citet{HL91} N$^M_{0.5}$ values to our $B_{gg}$ to be in the form
of a multiplicative constant.  \citet{HL91} account for luminosity
evolution by using the radio galaxy magnitude to define an epoch-invariant
point on the luminosity function and then referencing the magnitudes
of all galaxies in the counting region of the field to that of the
radio galaxy.  They note that this method yields values that are 
$\sim$ 20\% higher than
those from the method that we have adopted.  Thus, the conversion
constant might actually be a function of redshift.  However, the
objects in \citet{HL91} that are in our sample have a relatively
narrow redshift range ($z=0.367$ to $z=0.5524$) and so we assume a
single multiplicative constant for the conversion of all values
regardless of redshift.  Although we had no data in common
with \citet{HL91}, we had 1 $B_{gg}$ value from YE93 and 10
converted $B_{gg}$ values from \citet{Z97} that overlapped with the
\citet{HL91} data.  A weighted least squares linear fit forced through
the point (0,0) was performed using all 11 of these objects, resulting
in the conversion $B_{gg}$ = 33N$^M_{0.5}$.  The fit and the data points
with their error bars are shown in Figure 3.  
This conversion lies between the empirical conversion
($B_{gg}$ = 30N$_{0.5}$) and the theoretical conversion ($B_{gg}$ =
34N$_{0.5}$) given in \citet{HL91}.
Also, the somewhat lower conversion factor of the \citet{HL91} N$^M_{0.5}$
values as compared to that of the \citet{Z97} N$^{-19}_{0.5}$ values
(33 vs. 38) is consistent with the relationship between these values
quoted in \citet{AEZO93}, to within the error bars.

\subsection{Summary of All Clustering Data }
A compilation of the $B_{gg}$ values from all the methods discussed above
is given in Table 6.  Discrepancies and uncertainties are discussed in the
notes following the table and a key to the references is given after the
notes.  
Errors for our values were
discussed at the beginning of Section \ref{bgg}.  Errors for the YE93
values were taken directly from the literature.  Errors for the converted
$B_{gg}$ values (those from \citet{Z97}, \citet{HL91} and \citet{YMP89}) were
computed by converting the original $1\sigma$ values
to $B_{gg}$ values and using this spread in the converted $B_{gg}$ value as
the error.

``Adopted'' $B_{gg}$ values are listed in Table 7.  Sixty five of the 79 objects
in this complete 3CR sample have $B_{gg}$ values.  Because the absence of $B_{gg}$
data is due entirely to the absence of photometric images and because the order in
which objects were observed was random, there is no bias introduced due to this
incompleteness in the $B_{gg}$ data.  
In the absence of our own value, the value from YE93 was used
in Table 7, if available; otherwise converted $B_{gg}$ values were used.  
Since we had no data in common with \citet{HL91},
this conversion is indirect and so the values of \citet{Z97} were judged to generally
be the most reliable of the converted values, followed by those of \citet{HL91} and
then those of \citet{YMP89}.  Different converted values for the same field that
agreed to within their errors were averaged together.  
Two objects (3C\,275
and 3C\,435A) have a range of $B_{gg}$ values listed in Table 7 because other studies
yielded highly discrepant values and, in the absence of other data, it is unclear
which of the values is more reliable in these specific cases.

The distribution of the $B_{gg}$ values listed in Table 7 is shown in Figure 4.
The two objects (3C\,275 and 3C\,435A) with a range of $B_{gg}$ values listed in
Table 7 are omitted from Figure 4 and from all further analysis and discussions.
Table 7 contains 17 sources with $B_{gg} > 500$ Mpc$^{1.77}$.  The richness
of all 17 of these fields 
has been previously noted in one form or another in the literature.  Most notably,
3C\,28 has the richest environment (excluding the upper limit
on the range for 3C\,435A) and is a known Abell cluster (Abell 115) and a known
X-ray cluster (\citealt{MSv} and references therein).
In addition to the references listed in Tables 6 and 7, information on the fields
surrounding the remaining  sources with $B_{gg} > 500$ Mpc$^{1.77}$ can be
found in \citet{Wynd}, Kristian, Sandage \& Katem (1974, 1978), 
\citet{HinStoc}, \citet{E91},
\citet{Spinrad} and references therein, \citet{E94},
\citet{MSv}, \citet{Hall}, \citet{Travis}, \citet{Hes}
and \citet{GizLea}.

\section{Results}
\subsection{Radio Galaxy Environments vs. Quasar Environments}
The environments of the two samples must be compared as a function of
redshift because EYG have shown that quasar environments vary
dramatically with redshift.
Figure 5 shows the sample 
plotted both as a function of redshift and
absolute nuclear B magnitude.  Sources located in rich
environments ($B_{gg} > 500$ Mpc$^{1.77}$) are denoted by filled symbols
while those in poor environments ($B_{gg} \le 500$ Mpc$^{1.77}$) are marked
with open symbols.  
The absolute nuclear
B magnitude is in general agreement with the AGN classification; i.e.,
all quasars have $M_B < -22.5$ while all but 4 of the radio and N-galaxies
have $M_B > -22.5$.
For consistency,
the 4 galaxies with $M_B < -22.5$ are reclassified as quasars for
the environment comparisons discussed below (this has no significant
effect on the results).

Note the gap in magnitude between the quasars and most of the radio
galaxies in Figure 5.  It is unclear whether this gap is real or due to
possible systematic differences in the way the absolute nuclear B magnitude
was calculated for quasars and radio galaxies.  Quasars were assumed
to have a negligible host galaxy contribution and so a total magnitude was
used for the nuclear magnitude.  For radio galaxies, a magnitude in
the inner 1-2 kpc was used to estimate
the nuclear magnitude.
The systematic effect of including the host galaxy contribution in
the quasar nuclear magnitudes can be estimated using a ``typical''
AGN host galaxy magnitude of $M_{host_B} = -22$. (estimated from the
radio galaxy data in Table 3). 
Removing such a contribution makes an 
$M_B = -23$ quasar nuclear magnitude dimmer by 0.6 mag.
Thus, the net effect would be to spread out the lower envelope
of the quasar nuclear magnitude distribution,
but this is insufficient to close the gap in Figure 5 entirely.
So it seems possible that some systematic differences in magnitudes
could be present.
Still, this systematic is not large enough to cause a significant
overlap of quasar and radio galaxy absolute nuclear B magnitudes
(i.e., few objects would be misclassified).  

Also plotted in Figure 5 is a very low z ($z < 0.1$) sample of radio
galaxies, taken from YE93, which contains a
mixture of FR1 and FR2 type sources
(all those at $z <  0.1$  with  $B_{gg} > 500$ Mpc$^{1.77}$ 
are FR1s).  Although it has been included on this plot, this sample is NOT
included in any comparison between quasar and radio galaxy properties
because there are no quasars at this low redshift.
Originally, the case for an evolutionary fading model
for quasars was made by EYG on the basis of their quasar
sample at $z > 0.3$ and this very low z ($z < 0.1$) sample of radio galaxies.
The data from the present work fill the $\sim $ 4 Gyr gap 
between these two samples.

An inspection of just the quasars in Figure 5 shows that there are
a substantial number of quasars found in rich environments at
$0.4 < z < 0.65$ but there are none found in rich environments at
$z < 0.4$.  This dramatic result is presented and discussed in
EYG and YE93 and is the basis for their evolutionary
hypothesis.
We choose the $0.15 < z < 0.4$
redshift range in which to compare radio galaxy and quasar
environments because of the striking absence of quasars in rich
environments at these redshifts.
The orientation angle
hypothesis of \citet{Barthel} predicts that the percentage of
quasars found in rich environments at $0.15 < z < 0.4$ should
be the same as the percentage of radio galaxies found in rich
environments over that same redshift range.  
However, a Kolmogorov-Smirnov test yields only a 3\% chance
that the two distributions of B$_{gg}$ values  were drawn from
the same parent population (Figure 6).  On this basis the orientation angle
hypothesis of \citet{Barthel} is rejected at the 97\% (2.2$\sigma$)
confidence level.  

Since N-galaxies do have a strong point source component, it is
possible that N-galaxies are actually quasars with either a slightly lower
luminosity AGN or a slightly brighter host galaxy. 
If the above analysis is redone with {\it all}
N-galaxies grouped together with the quasars (rather than grouping
those with $M_B > -22.5$ with the radio galaxies), the
difference in the environments of the radio galaxies and quasars at
$0.15 < z < 0.4$ becomes even larger: a KS test gives only a 0.4\%
chance, or a rejection at the 3$\sigma$ level, that the two
distributions come from the same parent population. 

Because the AGN nuclear magnitudes are drawn from a continuous 
luminosity function with a somewhat unclear observational 
magnitude limit, we follow YE93 and quantify the relative evolution of rich and
poor environments by building a model of the brightest
AGN one would expect to see, as a function of redshift. 
We use the two-power-law model of the quasar luminosity function
from  \citet{Boyle}, with evolution in the
characteristic quasar luminosity parametrized as
L$^*(z) = L_0 (1+z)^{\kappa_L}$.  This luminosity function is used
along with the relation for cosmological volume as a function
of redshift, to produce a family of curves describing the
brightest AGN one would see at any given redshift, for various values of
the rate of evolution, $\kappa_L$. The curves are normalized
empirically to $z \sim 0.6$, where the brightest radio-loud quasars
observed in the 3CR sample are often found in rich environments.

In agreement with YE93, we find the ``standard
model'' ($\kappa_L = 3.7$; solid line) to be a good fit to the upper envelope
of all the data in Figure 5.  This model is based on the well-determined
parameters for optically-selected quasars, and hence describes the
general population of very luminous AGN, which here
are mostly located in poor environments.
The model corresponding to $\kappa_L = 19$ (dash-dot line) was found to be the best
match for the upper envelope of the data when only the sources in
rich environments in Figure 5 (i.e., only the filled symbols) were
used.  

An upper limit on the e-folding time ($\tau$) for the fading of quasars
can be obtained by assuming an exponential form for
the luminosity evolution, $L^*(z) = L_0e^{t/\tau}$.
For sources in rich ($B_{gg} > 500$ Mpc$^{1.77}$) environments
(i.e., $\kappa_L = 19$) at $z = 0.4$ (the middle of our redshift range),
the resulting e-folding time for quasars is
0.9 Gyr for $H_0 = 50, q_0 = 0$ (0.6 Gyr for $H_0 = 75, q_0 = 0$). 
Comparing the $\kappa_L$
values of these fits to the upper envelope of the sources in poor
($\kappa_L = 3.7$) and rich ($\kappa_L = 19$) environments, it
appears that sources in rich environments evolve $\sim 5$ times
faster than those in poor environments.
Changing the $B_{gg}$ value that defines a rich environment from 500 to
400 or 600 Mpc$^{1.77}$ has little effect on these results.
The upper envelope of the sources with $B_{gg} > 400$ Mpc$^{1.77}$ is
well fit by the model corresponding to $\kappa_L = 16$ while that of the
sources with $B_{gg} > 600$ Mpc$^{1.77}$ is well fit by a $\kappa_L = 24$
curve.
The e-folding times associated with these values of $\kappa_L$
vary by only $\sim 20\%$ (a 19\% increase in $\tau$ for $\kappa_L = 16$
and a 21\% decrease in $\tau$ for $\kappa_L = 24$) from those corresponding
to the $\kappa_L = 19$ curve.  
While changing
the definition of a rich environment to $B_{gg} > 700$ Mpc$^{1.77}$ still
has little effect on the results (the $\kappa_L = 24$ model is still a
good fit to this upper envelope), dropping the limit to $B_{gg} > 300$
Mpc$^{1.77}$ has a drastic effect.  In this case, the standard envelope
model ($\kappa_L = 3.7$) is a good fit and the e-folding times increase
by more than a factor of 5.

Although the evidence presented above 
makes it very unlikely 
that
the orientation angle hypothesis of \citet{Barthel} is the primary means
of relating quasars and radio galaxies, it is possible that {\it some}
quasars are radio galaxies with their jet oriented close to our line
of sight (within $\sim 45 ^\circ$).  Thus, Barthel's hypothesis
may be valid for individual objects but it is almost
certainly not valid for the AGN population as a whole.
Also, we cannot rule out that quasars and radio galaxies are related
to one another by some combination of both evolution and orientation.

\subsection{Correlations Between Environment and Other AGN Properties}
Since the 3CR sample is flux-limited, the radio power and redshift are
highly correlated ($r = 0.65$ at a confidence level $ > 99.9\%$).  Thus,
there is the possibility that the distribution of the objects in
rich environments seen in Figure 5 is actually due to a correlation between
radio power and environment at a given redshift combined with the radio
power-redshift correlation of the 3CR sample.
\citet{YMP89} do find a slight correlation between radio power and
environment in their study of radio galaxies, as do \citet{Wold}.
\citet{HL91}, whose
study of radio galaxies is well-suited to answer this question because of
its large range of radio powers and its relatively narrow redshift range,
find no evidence of a correlation between radio power and environment
at $z \sim 0.5$.  

In Figure 7, the total radio power at 178 MHz of all the 3CR objects used
in this study (including the radio galaxies with $z < 0.1$ taken from
YE93 which are not part of the present sample) are plotted
versus redshift with filled symbols indicating objects with
$B_{gg} > 500$ Mpc$^{1.77}$.  
The symbols have the same meaning as in Figure 5 and quasars
with unknown environments are marked with an ``X''.
A visual inspection of Figure 7 shows
that, with a few exceptions, for each object in a rich environment there
are objects at a similar redshift in a poor environment with equal or
greater radio power (i.e., the objects in rich environments, in general,
are not the most powerful radio sources at a given redshift).
Ignoring
the objects with $z < 0.1$, a correlation analysis
using the remaining 3CR objects (i.e., those in the sample of this work)
yields a correlation coefficient between radio power and environment of
$r = 0.27$ at a confidence level of $\sim 97\%$.
Thus, there does
appear to be a slight correlation between radio power and environment
for the objects in the present sample but not as strong a correlation
as would be required to create the apparent evolutionary effect seen
in Figure 5.  

Figure 8 shows the complete sample, including the quasars which are
not 3CR sources.
While a number of these objects lie at somewhat lower
radio powers than the radio galaxies at similar redshifts,
their powers are still well above many radio galaxies which
are located in clusters. The lack of clusters in this sample
is thus difficult to explain in a unified model.
We note that about 10\% of the quasar sample may be core dominated
(\citealt{Travis}, \citealt{hutchdata}), indicating that their total radio
powers may be overestimates of their lobe power. However, these objects
are found in both rich and poor environments, suggesting that 
this does not strongly affect the comparison of radio galaxies
and quasars.

Additionally, there is no significant correlation between total M$_r$
(and thus host galaxy M$_r$) and B$_{gg}$, although virtually all of these FR2
host galaxies are much more luminous than M$^*$. Specifically, eight of
the nine FR2s with M$_r$ determined and B$_{gg} \geq$ 500 Mpc$^{1.77}$ 
have absolute host
luminosities 1-2 mags brighter than M$^*$ (3CR 435.0A is the
lone exception with a poorly determined magnitude 0.5 mags less luminous
than M$^*$). These values are typical of poor cluster brightest cluster
galaxies (\citealt{Hoessel}; \citealt{Wurtz}), and, as found
by other studies (\citealt{Lilly}; \citealt{AEZO93}), 
the radio galaxies are the brightest members of their groups and poor
clusters.

Finally, we briefly examined the relationship between the
environment and the optical class (column 9 of Table 1) of each
source.  The distribution of $B_{gg}$ values was found to be similar
for all optical classes and no significant difference in the percentage
of rich environments ($B_{gg} > 500$ Mpc$^{1.77}$) for each optical
class was found.   
So, based on the correlation analyses
discussed above, the behavior illustrated in Figure 5 appears
to be genuine and not due to secondary and/or multiple
correlations of the various other AGN parameters.

\section{Conclusion}
\subsection{Summary of the Results}
The results of this work are summarized below:

1.  A comparison of the environments of 51 radio galaxies and 67
quasars in the redshift range $0.15 < z < 0.65$ clearly shows that
while quasars are found in rich environments ($B_{gg} > 500$ Mpc$^{1.77}$;
i.e., Abell richness class 0 and above)
only at $z > 0.4$ (EYG), radio galaxies are found in rich
environments over the entire redshift range with the percentage of radio
galaxies in rich environments at $0.15 < z < 0.4$ (20\%) being comparable
to that of radio galaxies in rich environments at $0.4 < z < 0.65$ (28\%)
and quasars in rich environments at $0.4 < z < 0.65$ (36\%).  A K-S test
gives only a 3\% chance that the $B_{gg}$ distributions of the radio
galaxies and quasars at $0.15 < z < 0.4$ come from the same parent
population.  On this basis, the orientation angle hypothesis of
\citet{Barthel} is rejected at the 97\% (2.2$\sigma$) confidence
level as the primary means of relating quasars and radio galaxies.
If all N-galaxies are grouped with the quasars, the hypothesis of
\citet{Barthel} is rejected at the 99.6\% (3$\sigma$) confidence level.
However, a combination of Barthel's orientation hypothesis and EYG's
evolutionary hypothesis cannot be ruled out.

2.  Applying the evolutionary model (pure luminosity evolution; i.e.,
no number density evolution) described in YE93, we find that
the ``standard envelope model'' ($\kappa_L = 3.7$) is a good fit to
the upper envelope of the sources in poor environments while a model
corresponding to $\kappa_L = 19$ matches the upper envelope of the
sources in rich ($B_{gg} > 500$ Mpc$^{1.77}$) environments.  A
comparison of these two values of $\kappa_L$ implies that sources
in rich environments evolve $\sim 5$ times faster than those in poor
environments.  Converting $\kappa_L$ to an e-folding time ($\tau$),
the maximum e-folding time for quasars in rich ($B_{gg} > 500$
Mpc$^{1.77}$) galaxy environments is
$\tau \sim 0.9$ Gyr for $H_0 = 50, q_0 = 0$ ($\tau \sim 0.6$ Gyr for
$H_0 = 75, q_0 = 0$).  All these results are in excellent agreement
with YE93 and are not strongly dependent upon the
exact choice of $B_{gg}$ used to define a rich environment.

3.  It is unlikely that the redshift relationship between the radio
galaxies in rich environments and the quasars in rich environments
seen in Figure 5 is due to a correlation between radio power and
environment coupled with the redshift-radio power correlation of
the flux-limited 3CR sample.  Nor do any other secondary or
multiple correlations appear to be responsible.

\subsection{A Compatible Hypothesis}

In this Section we describe a compatible hypothesis for the results presented here.
This hypothesis links the observed evolution of AGN activity in clusters
with the decline of individual sources in response to evolution in their
cluster environment.
While this may not be the only physical mechanism that could account for these results
(e.g., systematic changes in jet opening angle with time for AGN in clusters is another
possibility), the scenario we propose is simple and consistent with both theoretical ideas about
the triggering of quasars and observational constraints on the evolution of the cluster
gravitational potential well from X-ray observations. The ideas presented below also are most
easily understood in a model by which a single cluster AGN undergoes secular evolution in
power, instead of a statistical fading of cluster AGN as a population. 
While either possibility is consistent
with current data, we note that cD galaxies in rich clusters in the current epoch are
luminous radio emitters (i.e., typically log P$_{rad}\geq$25 WHz$^{-1}$) 35-40\% of the
time.
So in the case of the cD galaxies, like those in which the quasars and radio galaxies in
this study are located, a large fraction of the full population of such objects is radio-loud
at any one time, unless the current epoch statistics are very anomalous.
Since several lines of argument suggest that single radio source lifetimes
are 10$^{7-8}$ yrs, every cD galaxy must have ``turned on'' several times in the last few Gyrs
to account for the current epoch statistics. 
Therefore, particularly for the cD galaxies, a recurrent outburst hypothesis is particularly
attractive, and so we will assume that this idea is correct in the discussion below.

The triggering of quasars is believed to be caused by galaxy-galaxy
interactions and mergers (e.g., \citealt{Roos}; \citealt{BarnHern}; \citealt{WC95}).
However, the rate and relative velocities
of these interactions and mergers change as the clustering environment
around the quasar evolves.  In the early stages of cluster formation,
the duration of galaxy-galaxy interactions tends to be longer
because the cluster potential well is still shallow and the galaxies
have not yet reached their full orbital speed.
In this stage of cluster evolution, mergers are still
efficient at  destroying disk structures and transferring gas (e.g., 
\citealt{Mihos}). 
However, as a cluster virializes and begins to come into dynamical
equilibrium, the orbital speeds of the galaxies increase and the galaxy
orbits stabilize.  Galaxy-galaxy mergers and interactions become less
frequent and are of shorter duration, making them less effective at
triggering AGN activity (\citealt{Aarseth}; \citealt{Roos}; \citealt{WC95}).  

Several authors (e.g., \citealt{BZ77}; \citealt{BBR80};
\citealt{WC95}) have suggested
that, unlike radio-quiet AGN, radio-loud AGN are powered by the spin
energy extracted from very rapidly spinning supermassive black holes
(SMBHs). Rapidly spinning SMBHs can be formed most easily from the capture 
and subsequent orbital evolution of a SMBH binary system (on a timescale of
~10$^8$
yrs after the binary pair is formed; but see \citet{BBR80} who advocate
much longer timescales). Under certain circumstances, 
merging disk galaxies can provide non-rotating SMBHs from their nuclei 
(\citealt{WC95})
and the eventual merger of the SMBH pair 
forms the very rapidly spinning SMBH, whose
spin energy is extracted by the Blandford-Znajek mechanism 
to create the radio jets and other radio-loud quasar manifestations. 
As the cluster
virializes, the absence of suitably slow encounters prevents
the formation of new SMBH binaries as
well as the injection of mass into the region of the SMBH.  
Thus, the existing SMBH can only spin down as it releases
energy, creating lower power outbursts over time; i.e., a luminous quasar at
$z\sim$0.5 fades to become an FR2 radio galaxy then an FR1 radio galaxy.
In this picture, the fading timescale of 0.9 Gyrs found by this work
is the spin down time for the SMBH.

Alternately, the quasar is starved of fuel as the cluster virializes.
The nuclear engine of the quasar (i.e., the AGN) is believed to be
fueled by gas from either the quasar host galaxy or from an interacting
companion galaxy (\citealt{Roos}, 1985; \citealt{DeRob}).  In the early stages
of cluster formation, galaxies are believed to contain an abundance of
gas and so fuel for the AGN is plentiful.
However, as a cluster develops, an intracluster medium (ICM) begins to
form due to the combined remnants of galaxy-galaxy interactions and
mergers (e.g., tidal tails).  As the ICM becomes dense
($> 10^{-4} $cm$^{-3}$), it will begin to strip or ``sweep'' gas out
of the galaxies moving through it (\citealt{StocPerr}), making itself
more dense and even more effective at ``sweeping'' gas (\citealt{Gisler}).
As the gas sweeping ability of the ICM increases, potential fuel for
the AGN is more easily stripped from cluster galaxies moving through
the medium and heated to X-ray emitting temperatures.  This positive
feedback loop may eventually starve
the AGN (by decreasing the gas reservoir in both the AGN host galaxy
and the interacting companion galaxies) causing it to dim.

In this scenario, therefore, the evolution of the fraction of AGN seen in clusters 
probes the systematic evolution of the physical conditions within the cluster
environment. Because of their sensitivity to local
conditions, the most optically luminous AGN mark primarily
poor groups {\it or} those richer systems which are in the early stages of
virilization.
Thus, clusters of a given richness will
no longer harbor quasars past a certain epoch.  For clusters having a
richness of $B_{gg} \sim 500$ Mpc$^{1.77}$, this epoch is $z \sim 0.4$
(EYG).  
This quasar dimming will occur at an earlier epoch for
even richer environments than those discovered here (i.e., Abell richness
class 2 and higher) while for poorer environments it will occur closer
to the present time.
That is, to our knowledge, no quasar or luminous radio galaxy (i.e., FR2)
has been discovered in an {\it extremely} rich cluster
($B_{gg} > 1500$ Mpc$^{1.77}$) at $z < 0.65$.  The scenario described here
suggests that this absence occurs because the ICM in extremely rich clusters is
already too dense and quasars therein have been ``starved'' at earlier
epochs.  Thus, by $z \sim 0.6$ these AGN have faded through luminous
FR2 radio galaxies to become low power FR1 radio galaxies.  The study
of radio sources in the richest cluster environments
($B_{gg} > 1500$ Mpc$^{1.77}$) at $z = 0.3 - 0.8$ by \citet{chap5}
supports this suggestion, finding only FR1 type sources in these clusters.

The maximum evolutionary e-folding time of $\sim 0.9$ Gyr is
a factor of several shorter than the virialization timescale for
the entire cluster; however, quasar evolution is more likely to be affected 
by the conditions in the cluster core region, rather than the 1 Mpc
scale environment.
Thus, even if cluster virialization triggers the fading of these
quasars, the fading itself occurs too rapidly to be
``tracking'' the subsequent cluster evolution.  However, evolution
of the cluster core could
``track'' quasar fading.  
Assuming a cluster core radius of
0.1 - 0.25 Mpc (based on the extent of X-ray emission from clusters;
Henry et al.\ 1992) and 5 crossing times for virialization, the
virialization time of a cluster core would be $\sim 1-2.5$ Gyr
which is more comparable to the quasar fading time we find above.
Therefore, while cluster virialization initially "starves" the quasar by
one or both of the mechanisms above, the actual fading timescale must be
due to some internal AGN clock, not to any timescale in the larger
environment of the AGN, all of which are too long.

\subsection{Further Tests of This Hypothesis}
Additional studies can test some of the ideas proposed in this work.
Based on the scenario presented above, clusters of richness
$B_{gg} \gtrsim 500$ Mpc$^{1.77}$ at $z > 0.4$ hosting AGN would be expected to
have lower velocity dispersions  and higher fractions of blue,
star-forming galaxies
than clusters of similar richness at
$z < 0.4$ (e.g., \citealt{E94}). 
The link between ICM and AGN activity would
suggest that AGN-hosting clusters of richness
$B_{gg} \gtrsim 500$ Mpc$^{1.77}$ at $z < 0.4$ have a denser ICM than
clusters of similar richness at $z > 0.4$; \citet{Hall}
show that extended X-ray luminosities around quasars fall in the
lower range of what is expected from clusters of similar richness.
A more conclusive test of this
hypothesis could be performed using sensitive, high resolution
X-ray images such as those from ACIS aboard Chandra to search for
direct detections of a dense, hot ICM surrounding both the quasars and 
radio galaxies at $z < 0.4$.
An indirect investigation
of the ICM surrounding the sources studied here using the extended
radio morphology of these sources is presented in Paper 2; the
results support this hypothesis and suggest that a dense ICM can
affect extended radio structure and may even cause FR2 type radio
sources to evolve into FR1-like structures at $z < 0.4$.  

\acknowledgments

We wish to thank KPNO for telescope
time and support given to us for this project.
M.H. acknowledges the support of a NASA Graduate Student Research
Program Fellowship NGT - 51291 and the travel support of NOAO to
conduct these thesis observations. EE would like to acknowledge
NSF grant AST-9617145. H.K.C. Yee is thanked for the
use of his PPP program and for numerous helpful discussions.
G. Hill is also thanked for his cooperation.  This research has made
use of the NASA/IPAC Extragalactic Database (NED) which is operated
by the Jet Propulsion Laboratory, California Institute of Technology,
under contract with NASA.

\clearpage

%
%

%

\clearpage

\begin{figure}
\hoffset -500pt
\plotfiddle{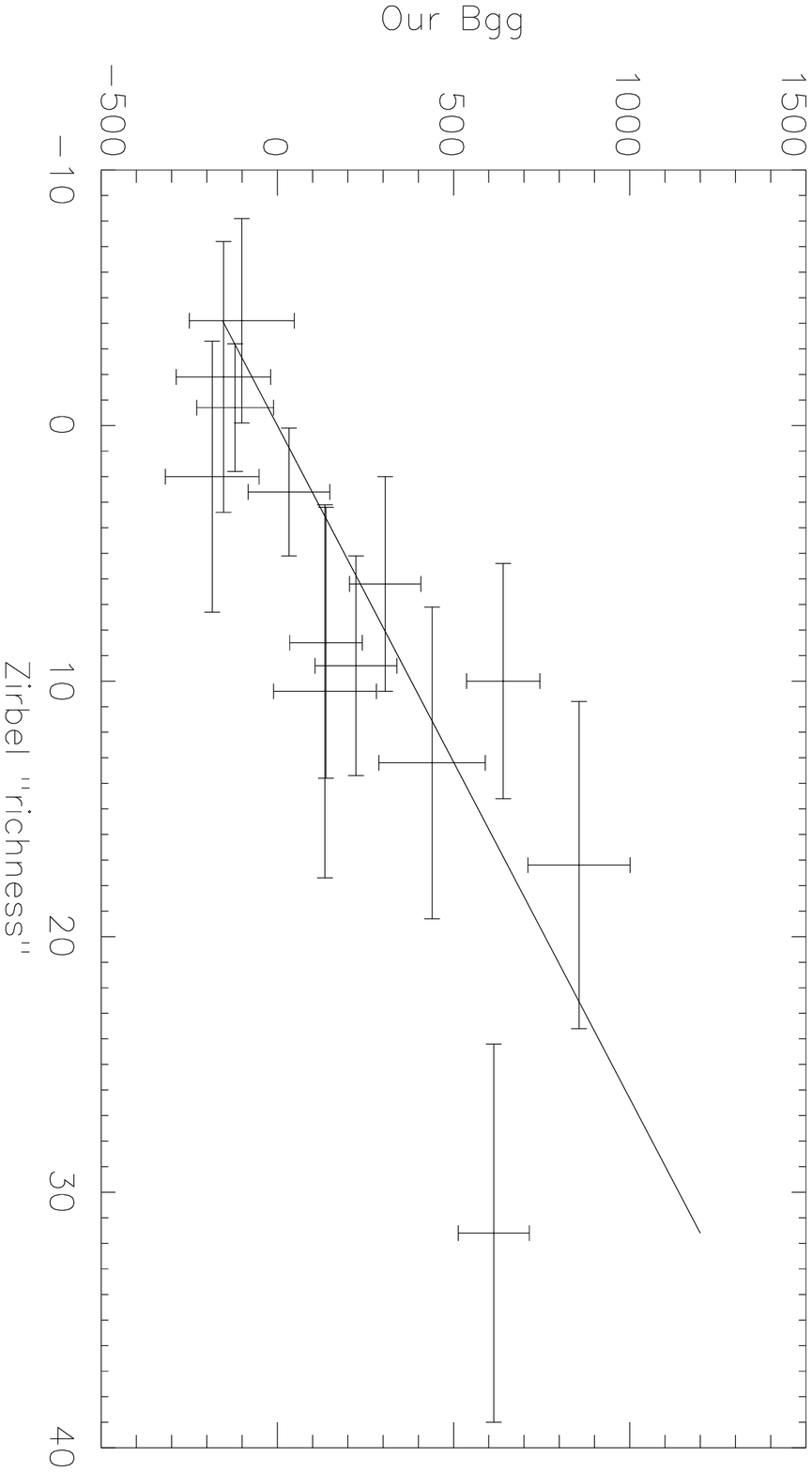}{1.0truein}{90}{0.5}{0.5}{-10000.}{0}
\caption{The data points with $1\sigma$ error bars and the
weighted linear fit that is the conversion from Zirbel (1997)
``richness'' values (N$^{-19}_{0.5}$) to $B_{gg}$.  The fit has
been forced through the point (0,0).  The resulting conversion is
$B_{gg}$ = 38N$^{-19}_{0.5}$.}
\end{figure}

\begin{figure}
\plotfiddle{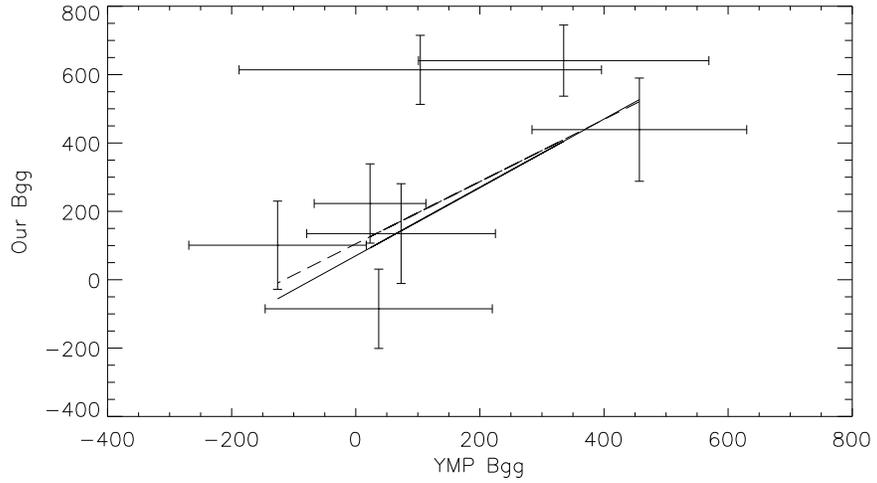}{1.0truein}{90}{0.5}{0.5}{-500}{50}
\caption{The data points with $1\sigma$ error bars and two
possible fits for the conversion from
Yates, Miller \& Peacock (1989; YMP) $B_{gg}$ values to our
$B_{gg}$ values.  The solid line is the conversion based on the
average offsets of the data points (i.e., a forced slope of unity).
The dashed line is a weighted least squares linear fit.  The two fits
agree to within their errors.  The two uppermost data points
were not used in the calculation of either fit (for reasons given
in the text).  The average offset fit (the solid line;
$B_{gg}$ = YMP $B_{gg}$ + 70) was used for the conversion.}
\end{figure}

\begin{figure}
\plotfiddle{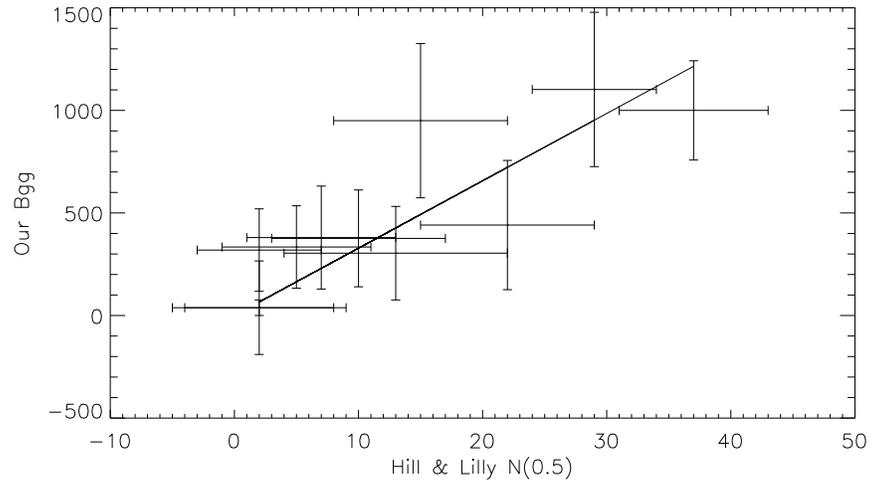}{1.0truein}{90}{0.5}{0.5}{235}{50}
\caption{The data points with $1\sigma$ error bars
and the weighted linear fit that is the conversion from Hill \& Lilly
(1991; HL) N$^M_{0.5}$ values to $B_{gg}$.  The fit has been forced
through the point (0,0).  The resulting conversion is
$B_{gg}$ = 33N$^M_{0.5}$.
This fit agrees quite well with the conversions discussed in HL.}
\end{figure}

\begin{figure}
\plotfiddle{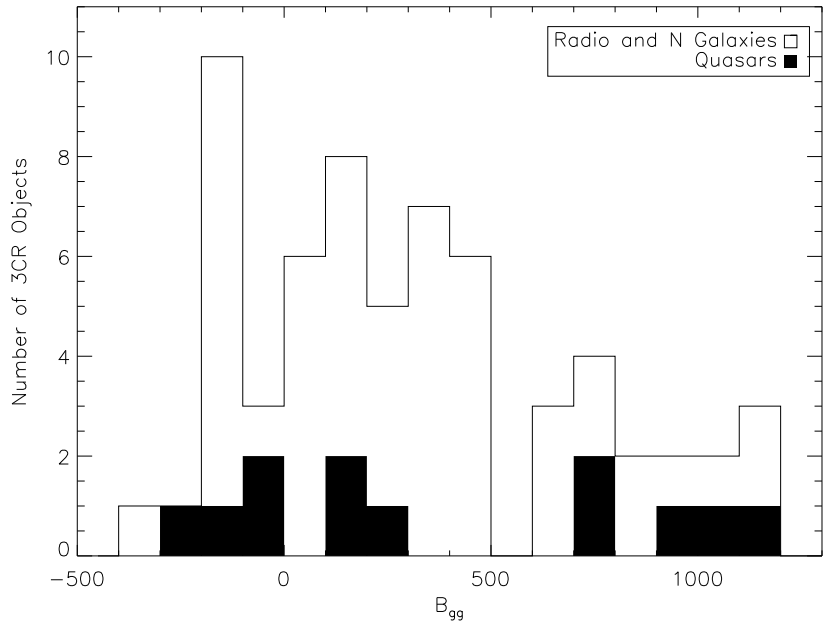}{1.0truein}{90}{0.5}{0.5}{252}{0}
\caption{A histogram showing the distribution of $B_{gg}$ values.
The open blocks are the radio galaxies and N-galaxies; the shaded
blocks are the quasars.  The two objects with a very large
discrepancy between $B_{gg}$ values found by other investigations (3C\,275
and 3C\,435A) are not included in the distribution.}
\end{figure}

\begin{figure}
\plotfiddle{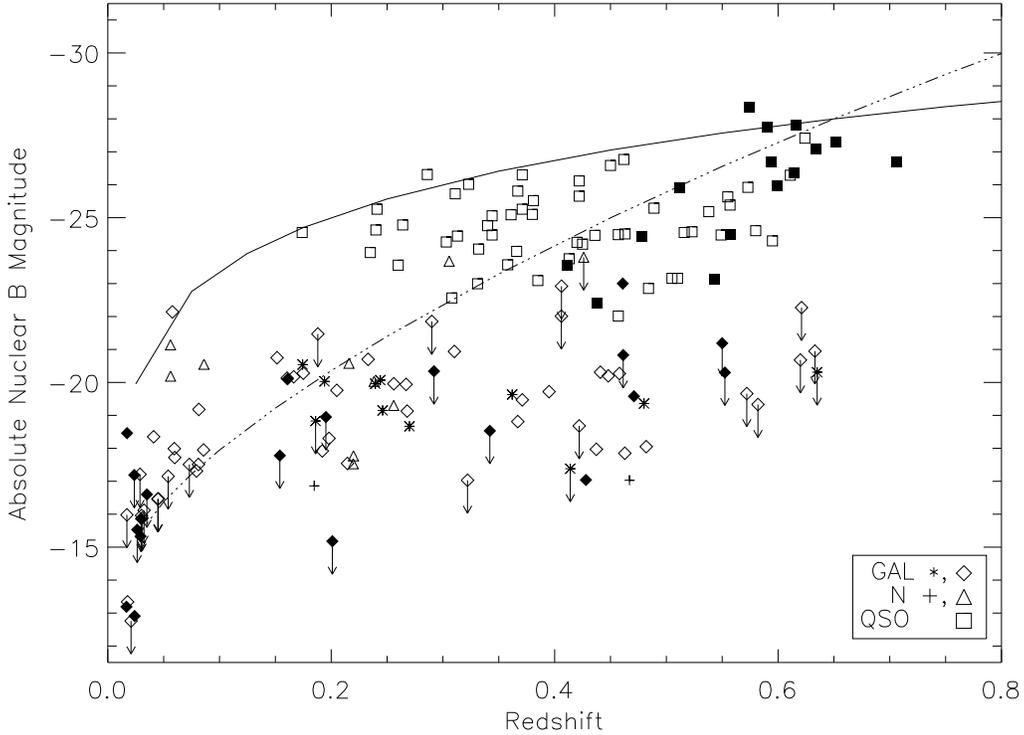}{1.0truein}{90}{0.6}{0.6}{245}{0}
\caption{Nuclear $M_B$ vs. $z$ for radio galaxies, N-galaxies and
quasars.  The absolute magnitudes are in rest-frame Johnson B and have
been corrected for Galactic reddening.  The radio galaxies appear as
diamonds, the N-galaxies as triangles and the quasars as squares.
Sources that are located in rich environments
($B_{gg} > 500$ Mpc$^{1.77}$) are denoted by filled symbols while
those in poor environments ($B_{gg} \le 500$ Mpc$^{1.77}$) are
marked with open symbols.  For some sources the environment is unknown
and these are marked with either an asterisk (radio galaxies) or
a plus sign (N-galaxies).  Symbols with an arrow indicate that the
absolute nuclear B magnitude is an upper limit.  The $z = 0.15 - 0.65$
radio and N-galaxies are from the present work, the lower redshift
galaxies are from Yee \& Ellingson (1993) and the quasars are
originally from Ellingson, Yee \& Green (1991; EYG).
Model curves for the upper envelope of the points are derived from
an  AGN luminosity function, using a 
``standard $\kappa_L=3.7$ model" (solid line), matching the upper
envelope for all objects,
and a $\kappa_L = 19$ model (dot-dashed line) 
describing the
upper envelope for the quasars and radio galaxies in 
rich environments (see text for details).
}
\end{figure}

\begin{figure}
\plotfiddle{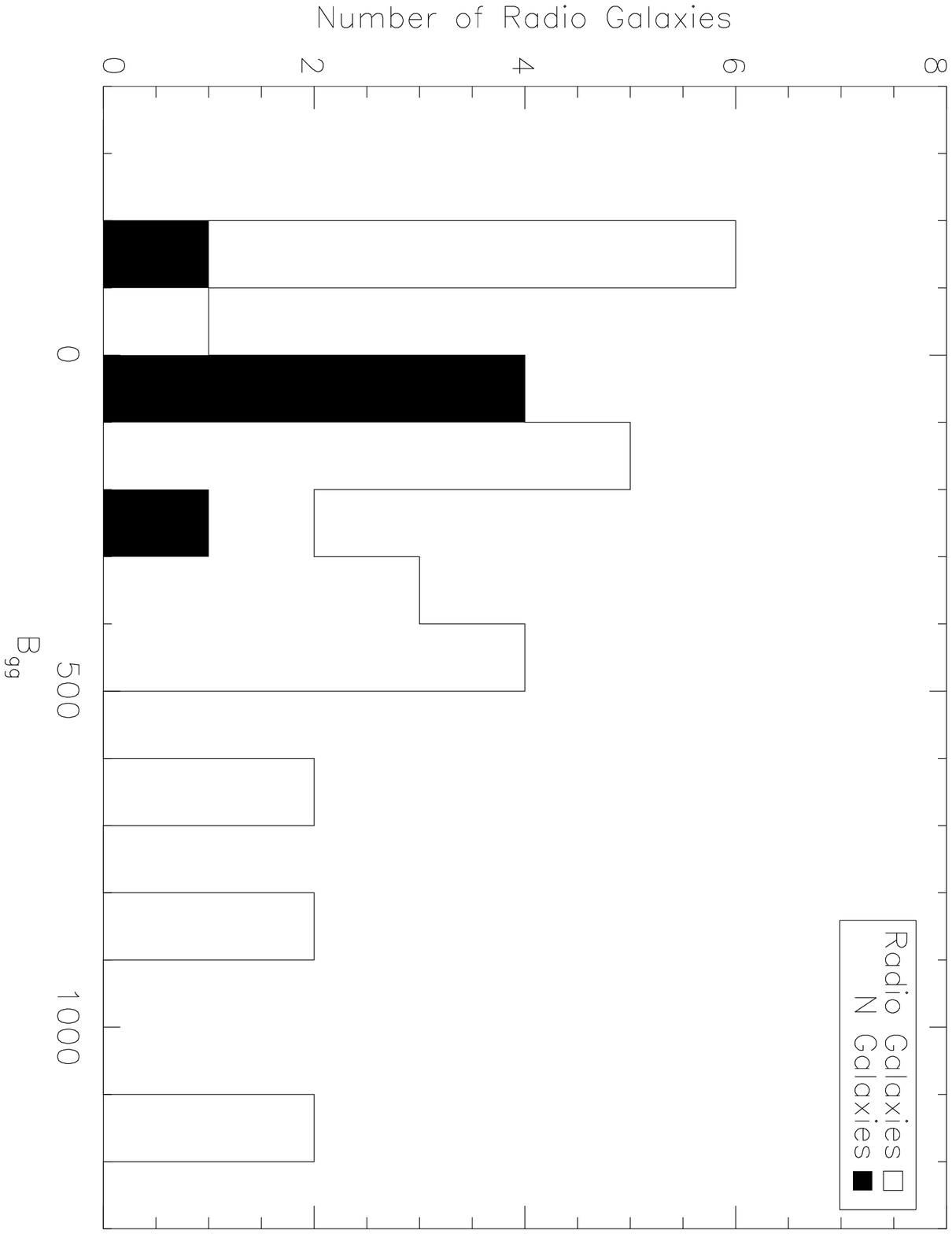}{1,0in}{90}{0.5}{0.5}{180}{0}
\plotfiddle{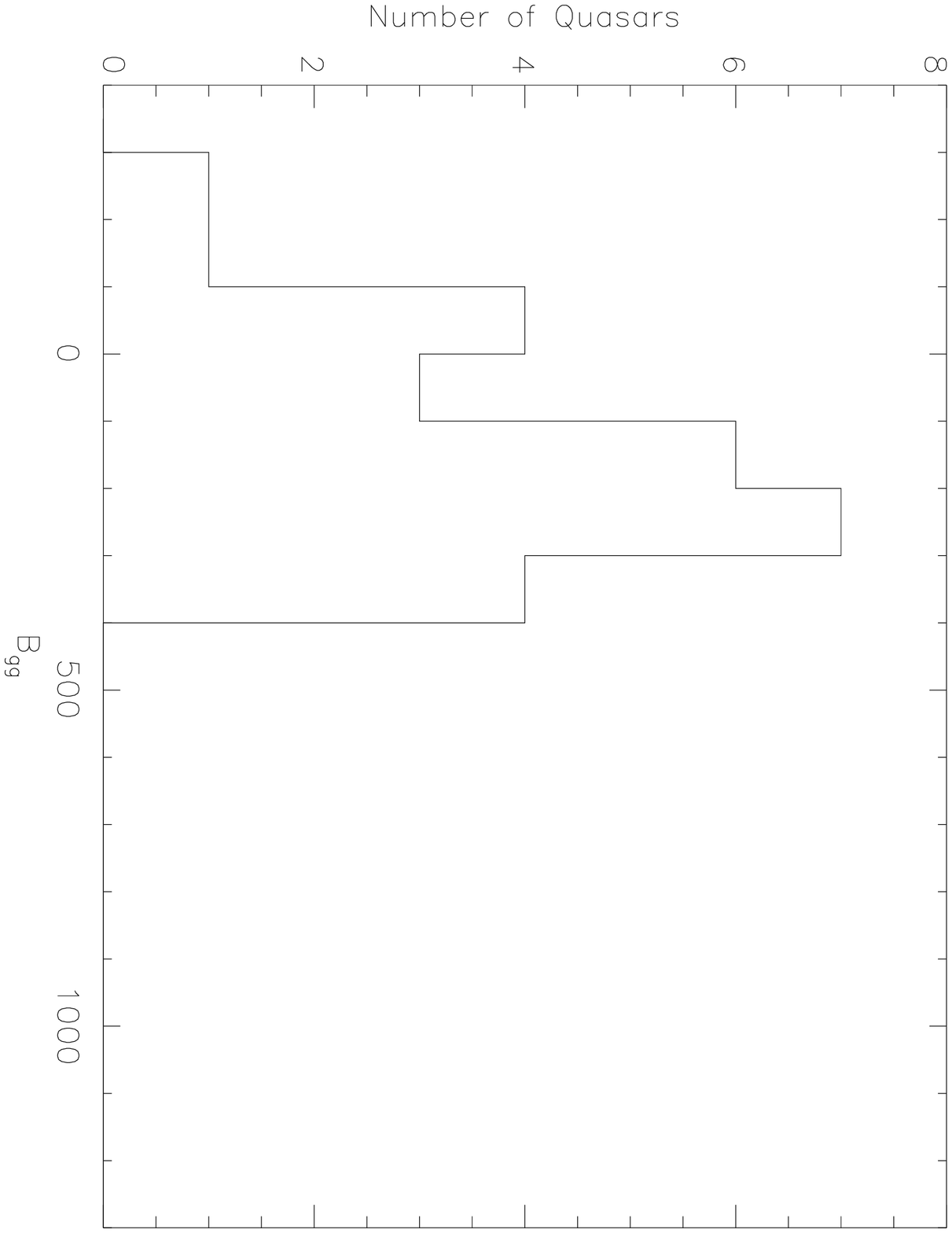}{1.0in}{90}{0.5}{0.5}{180}{0}
\caption{ a)The distribution of $B_{gg}$ values
for the galaxy environments of radio galaxies and N-galaxies
with $0.15 < z < 0.4$.  The N-galaxies are the shaded blocks.
b) The distribution of $B_{gg}$ values for the galaxy
environments of quasars with $0.15 < z < 0.4$.  A KS-test gives
only a 3\% chance that these two distributions come from the
same parent population. If the N-galaxies are grouped with the quasars,
the probability is 0.4\%.
}
\end{figure}

\begin{figure}
\plotfiddle{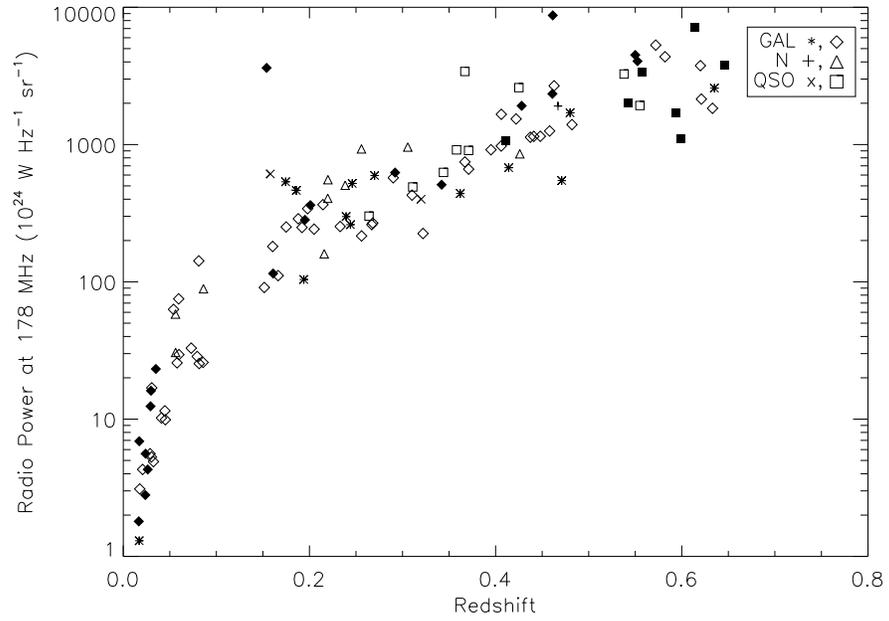}{1.0truein}{90}{0.5}{0.5}{244}{0}
\caption{Total $P_{178}$ vs. $z$ for all 3CR objects (both
galaxies and quasars) used in this study (including the radio
galaxies with $z < 0.1$ taken from Yee \& Ellingson 1993).  The
radio power is the total power emitted in the rest frame at 178 MHz.
The symbols have the same meaning as in Figure 5.  Two additional
quasars with unknown environments are marked with an ``X''.}
\end{figure}

\begin{figure}
\plotfiddle{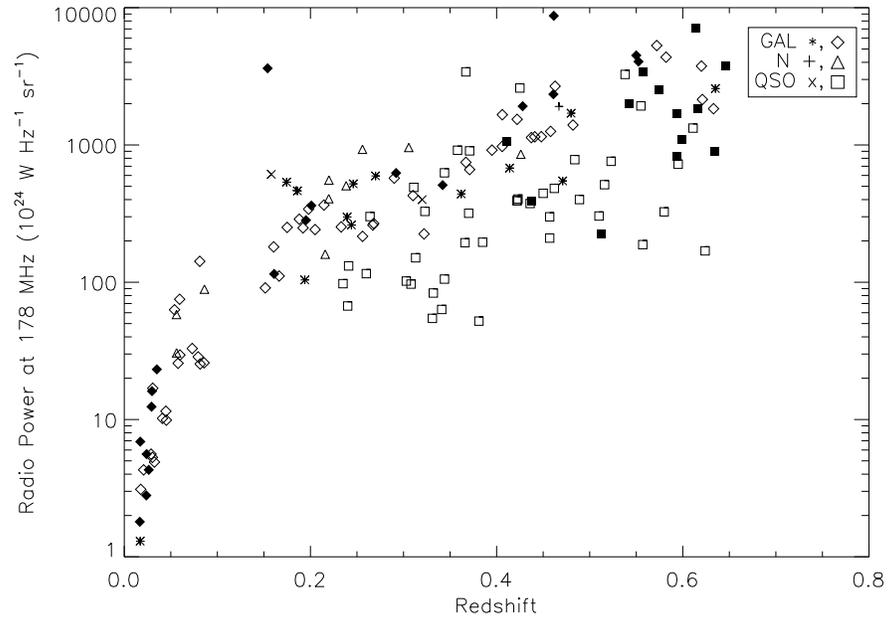}{1.0truein}{90}{0.5}{0.5}{244}{0}
\caption{ Same as Figure 7, but includes all objects in
the combined radio galaxy and quasar samples.}
\end{figure}

\clearpage 

\begin{table}
\tabletypesize{\scriptsize}
\caption{The Sample of 3CR Sources}
\label{sample}
\tiny
\begin{tabular}{cccccccccc}
\tableline \tableline
Source & R.A. & Dec. & $z$ & $S_{178}$ & $\alpha^{750}_{178}$ & $P_{178}$ (10$^{24}$ & log $P_{178}$ & Optical & ID \\
(3CR\#) & (B1950.) & (B1950.) & & (Jy) &  &  W Hz$^{-1}$ sr$^{-1}$) &   & Class &   \\ \tableline
    16.0  & 00 35 09.16 &  13 03 39.6  & 0.406  & 12.2  & 0.94   &  976  &   26.99   &  N  &  GAL \\
    17.0  & 00 35 47.18 & -02 24 09.5  & 0.2197 & 21.8  & 0.55   &  406  &   26.61   &  B  &  N   \\
    18.0  & 00 38 14.57 &  09 46 56.1  & 0.188  & 20.7  & 0.76   &  288  &   26.46   &  N  &  GAL \\
    19.0  & 00 38 13.76 &  32 53 39.9  & 0.482  & 13.2  & 0.63   & 1398  &   27.15   &  N? &  GAL \\
    28.0  & 00 53 09.12 &  26 08 23.4  & 0.1952 & 17.8  & 1.06   &  283  &   26.45   & --- &  GAL \\
    42.0  & 01 25 42.67 &  28 47 30.4  & 0.395  & 13.1  & 0.73   &  917  &   26.96   &  N  &  GAL \\
    46.0  & 01 32 34.09 &  37 38 47.0  & 0.4373 & 11.1  & 1.13   & 1131  &   27.05   &  N  &  GAL \\
    47.0  & 01 33 40.42 &  20 42 10.6  & 0.425  & 28.8  & 0.98   & 2600  &   27.41   &  B  &  QSO \\
    48.0  & 01 34 49.82 &  32 54 20.4  & 0.367  & 60.0  & 0.59   & 3409  &   27.53   & --- &  QSO \\
    49.0  & 01 38 28.41 &  13 38 19.9  & 0.621  & 11.2  & 0.65   & 2145  &   27.33   & --- &  GAL \\
    61.1  & 02 10 37.1  &  86 05 18.5  & 0.186  & 34.0  & 0.77   &  463  &   26.67   &  E  &  GAL \\
    63.0  & 02 18 21.90 & -02 10 33.   & 0.175  & 20.9  & 0.81   &  251  &   26.40   &  N  &  GAL \\
    67.0  & 02 21 18.05 &  27 36 37.4  & 0.3102 & 10.9  & 0.58   &  428  &   26.63   & --- &  GAL \\
    79.0  & 03 07 11.48 &  16 54 36.9  & 0.2559 & 33.2  & 0.92   &  930  &   26.97   &  N  &  N   \\
    93.0  & 03 40 51.54 &  04 48 21.7  & 0.358  & 15.7  & 0.85   &  915  &   26.96   &  B  &  QSO \\
    93.1  & 03 45 35.80 &  33 44 05.9  & 0.244  & 10.8  & 0.73   &  261  &   26.42   & --- &  GAL \\
    99.0  & 03 58 33.28 &  00 28 10.6  & 0.426  &  9.6  & 0.93   &  856  &   26.93   & --- &  N   \\
   109.0  & 04 10 54.85 &  11 04 39.5  & 0.3056 & 23.5  & 0.85   &  959  &   26.98   &  B  &  N   \\
   142.1  & 05 28 48.1  &  06 28 14.8  & 0.4061 & 21.1  & 0.89   & 1661  &   27.22   & --- &  GAL \\
   169.1  & 06 47 35.5  &  45 13 01.   & 0.633  &  8.0  & 0.93   & 1838  &   27.26   &  N  &  GAL \\
   171.0  & 06 51 11.05 &  54 12 50.0  & 0.2384 & 21.3  & 0.87   &  505  &   26.70   &  N  &  N   \\
   173.1  & 07 02 47.91 &  74 54 16.6  & 0.292  & 16.8  & 0.88   &  625  &   26.80   &  E  &  GAL \\
   196.1  & 08 12 57.32 & -02 59 13.9  & 0.198  & 20.3  & 1.19   &  341  &   26.53   & --- &  GAL \\
   200.0  & 08 24 21.43 &  29 28 42.2  & 0.458  & 12.3  & 0.84   & 1256  &   27.10   &  E  &  GAL \\
   213.1  & 08 58 05.15 &  29 13 34.5  & 0.194  &  7.2  & 0.58   &  104  &   26.02   &  N  &  GAL \\
   215.0  & 09 03 44.15 &  16 58 15.7  & 0.411  & 12.4  & 1.06   & 1064  &   27.03   &  B  &  QSO \\
   219.0  & 09 17 50.70 &  45 51 44.2  & 0.1744 & 44.9  & 0.81   &  536  &   26.73   &  B  &  GAL \\
   220.1  & 09 26 31.87 &  79 19 45.4  & 0.620  & 17.2  & 0.93   & 3756  &   27.57   &  N  &  GAL \\
   225.0B & 09 39 32.19 &  13 59 33.3  & 0.582  & 23.2  & 0.94   & 4362  &   27.64   &  N? &  GAL \\
   228.0  & 09 47 27.72 &  14 34 02.9  & 0.5524 & 23.8  & 1.00   & 4050  &   27.61   &  N  &  GAL \\
   234.0  & 09 58 57.38 &  29 01 37.4  & 0.1848 & 34.2  & 0.86   &  466  &   26.67   & N/B?&  N   \\
   244.1  & 10 30 19.61 &  58 30 04.3  & 0.428  & 22.1  & 0.82   & 1916  &   27.28   &  N  &  GAL \\
   249.1  & 11 00 27.42 &  77 15 08.7  & 0.311  & 11.7  & 0.81   &  491  &   26.69   &  B  &  QSO \\
   258.0  & 11 22 06.42 &  19 35 58.8  &        &  3.7  & 0.60   &       &           &     &      \\
   263.0  & 11 37 08.97 &  66 04 26.9  & 0.646  & 16.6  & 0.82   & 3795  &   27.58   &  B  &  QSO \\
   268.2  & 11 58 24.8  &  31 50 02.   & 0.362  &  7.5  & 0.79   &  440  &   26.64   &  N  &  GAL \\
   268.3  & 12 03 54.28 &  64 30 18.6  & 0.371  & 11.7  & 0.50   &  662  &   26.82   & --- &  GAL \\
   273.0  & 12 26 33.35 &  02 19 42.0  & 0.158  & 68.5  & 0.26   &  611  &   26.79   &  B  &  QSO \\
   274.1  & 12 32 56.74 &  21 37 05.8  & 0.422  & 18.0  & 0.87   & 1537  &   27.19   &  E? &  GAL \\
   275.0  & 12 39 45.16 & -04 29 53.9  & 0.480  & 15.8  & 0.70   & 1704  &   27.23   &  E  &  GAL \\
   275.1  & 12 41 27.58 &  16 39 18.0  & 0.557  & 19.9  & 0.96   & 3394  &   27.53   &  B  &  QSO \\
   277.0  & 12 49 26.15 &  50 50 42.9  & 0.414  &  8.2  & 0.91   &  679  &   26.83   &  E? &  GAL \\
   277.1  & 12 50 15.13 &  56 50 36.4  & 0.320  &  9.3  & 0.67   &  400  &   26.60   & --- &  QSO \\
   284.0  & 13 08 41.38 &  27 44 02.6  & 0.2394 & 12.3  & 0.95   &  299  &   26.48   &  N  &  GAL \\
   287.1  & 13 30 20.46 &  02 16 09.0  & 0.2159 &  8.9  & 0.55   &  160  &   26.20   &  B  &  N   \\
   288.0  & 13 36 38.59 &  39 06 21.8  & 0.246  & 20.6  & 0.85   &  521  &   26.72   & --- &  GAL \\
   295.0  & 14 09 33.44 &  52 26 13.6  & 0.4614 & 91.0  & 0.63   & 8730  &   27.94   & --- &  GAL \\
   299.0  & 14 19 06.29 &  41 58 30.2  & 0.367  & 12.9  & 0.65   &  747  &   26.87   & --- &  GAL \\
   300.0  & 14 20 40.10 &  19 49 13.2  & 0.270  & 19.5  & 0.78   &  595  &   26.77   &  N  &  GAL \\
   303.1  & 14 43 53.7  &  77 20 05.   & 0.267  &  8.8  & 0.77   &  261  &   26.42   & --- &  GAL \\
   306.1  & 14 52 24.5  & -04 08 47.   & 0.441  & 12.4  & 0.81   & 1147  &   27.06   &  N  &  GAL \\
   313.0  & 15 08 32.66 &  08 02 48.2  & 0.461  & 22.5  & 0.85   & 2342  &   27.37   &  N  &  GAL \\
   319.0  & 15 22 43.90 &  54 38 38.4  & 0.192  & 16.7  & 0.90   &  249  &   26.40   &  E  &  GAL \\
   320.0  & 15 29 29.70 &  35 43 48.5  & 0.342  &  9.9  & 0.78   &  510  &   26.71   &  E  &  GAL \\
   323.1  & 15 45 31.11 &  21 01 32.5  & 0.264  & 10.6  & 0.68   &  301  &   26.48   &  B  &  QSO \\
   327.1  & 16 02 12.96 &  01 25 58.7  & 0.4628 & 25.7  & 0.83   & 2679  &   27.43   &  N  &  GAL \\
   330.0  & 16 09 13.90 &  66 04 22.8  & 0.550  & 30.3  & 0.71   & 4492  &   27.65   &  N  &  GAL \\
   332.0  & 16 15 47.27 &  32 29 45.0  & 0.1515 & 10.5  & 0.64   &   91  &   25.96   &  B  &  GAL \\
   334.0  & 16 18 07.40 &  17 43 30.5  & 0.555  & 11.9  & 0.86   & 1925  &   27.28   &  B  &  QSO \\
   337.0  & 16 27 19.07 &  44 25 38.2  & 0.635  & 12.9  & 0.63   & 2577  &   27.41   &  N  &  GAL \\
   341.0  & 16 26 02.4  &  27 48 14.   & 0.448  & 11.8  & 0.85   & 1149  &   27.06   &  N  &  GAL \\
   345.0  & 16 41 17.60 &  39 54 10.7  & 0.594  & 11.8  & 0.27   & 1706  &   27.23   &  B  &  QSO \\
   346.0  & 16 41 34.56 &  17 21 20.7  & 0.161  & 11.9  & 0.52   &  115  &   26.06   & --- &  GAL \\
   348.0  & 16 48 39.98 &  05 04 35.0  & 0.154  &382.6  & 1.03   & 3619  &   27.56   &  E  &  GAL \\
   349.0  & 16 58 04.44 &  47 07 20.3  & 0.205  & 14.5  & 0.74   &  242  &   26.38   &  N  &  GAL \\
   351.0  & 17 04 03.51 &  60 48 31.3  & 0.371  & 14.9  & 0.73   &  906  &   26.96   &  B  &  QSO \\
   357.0  & 17 26 27.41 &  31 48 23.9  & 0.1664 & 10.6  & 0.59   &  111  &   26.04   &  N  &  GAL \\
   379.1  & 18 25 55.93 &  74 19 06.8  & 0.256  &  8.1  & 0.70   &  216  &   26.33   &  N  &  GAL \\
   381.0  & 18 32 24.40 &  47 24 36.5  & 0.1605 & 18.1  & 0.81   &  181  &   26.26   &  B  &  GAL \\
   401.0  & 19 39 38.84 &  60 34 32.6  & 0.201  & 22.8  & 0.71   &  362  &   26.56   &  E  &  GAL \\
   411.0  & 20 19 44.19 &  09 51 33.8  & 0.467  & 18.0  & 0.82   & 1909  &   27.28   &  N  &  N   \\
   427.1  & 21 04 44.80 &  76 21 09.5  & 0.572  & 29.0  & 0.97   & 5300  &   27.72   &  E  &  GAL \\
   434.0  & 21 20 54.40 &  15 35 11.7  & 0.322  &  5.2  & 0.64   &  225  &   26.35   &  E  &  GAL \\
   435.0A & 21 26 37.01 &  07 19 52.4  & 0.471  &  4.9  & 0.90   &  547  &   26.74   &  N  &  GAL \\
   436.0  & 21 41 57.91 &  27 56 30.3  & 0.2145 & 19.4  & 0.86   &  365  &   26.56   &  N  &  GAL \\
   455.0  & 22 52 34.53 &  12 57 33.5  & 0.5427 & 14.0  & 0.71   & 2012  &   27.30   & --- &  QSO \\
   456.0  & 23 09 56.65 &  09 03 07.8  & 0.2330 & 11.6  & 0.72   &  253  &   26.40   &  N  &  GAL \\
   458.0  & 23 10 21.9  &  05 00 26.   & 0.290  & 15.8  & 0.84   &  573  &   26.76   &  N  &  GAL \\
   459.0  & 23 14 02.27 &  03 48 55.2  & 0.2199 & 27.9  & 0.87   &  555  &   26.74   &  N  &  N   \\
   460.0  & 23 18 59.75 &  23 30 20.4  & 0.268  &  8.9  & 0.80   &  268  &   26.43   &  N  &  GAL \\ \tableline
\end{tabular}
\normalsize
\end{table}

\begin{table}
\tabletypesize{\scriptsize}
\caption{Optical Observations}
\label{optobs}
\tiny
\begin{tabular}{ccccccc}
\tableline \tableline
Source & Filter & UT Date & Telescope & Integration & Seeing & $5\sigma$ Limiting \\
(3CR\#) & & & & Time (sec) & (arcsec) & Magnitude \\ \tableline
    16.0  & r &    06NOV96 & 2.1 m & 1800 & 1.3 & 23.88 \\
          & g &    07NOV96 & 2.1 m & 3600 & 1.6 & 24.36 \\
    17.0  & r &    09NOV96 & 2.1 m &  900 & 0.9 & 23.58 \\
          & g &    09NOV96 & 2.1 m & 1800 & 1.2 & 24.09 \\
    18.0  & r &    08NOV96 & 2.1 m &  900 & 1.4 & 23.13 \\
          & g &    06NOV96 & 2.1 m & 1800 & 1.5 & 24.05 \\
    19.0  & r &    05NOV96 & 2.1 m & 1800 & 1.0 & 24.11 \\
          & g &    08NOV96 & 2.1 m & 3600 & 1.7 & 24.28 \\
    28.0  & r &    06NOV96 & 2.1 m &  900 & 1.5 & 23.29 \\
          & g &    08NOV96 & 2.1 m & 1800 & 1.4 & 23.88 \\
    42.0  & r &    05NOV96 & 2.1 m & 1800 & 0.9 & 24.20 \\
          & g &    08NOV96 & 2.1 m & 3600 & 1.4 & 24.45 \\
    46.0  & r &    09NOV96 & 2.1 m & 1800 & 0.9 & 24.03 \\
    47.0  & r &    08NOV96 & 2.1 m & 1800 & 1.4 & 23.64 \\
          & g &    09NOV96 & 2.1 m & 3600 & 1.1 & 24.56 \\
    49.0  & r &    05NOV96 & 2.1 m & 2700 & 0.8 & 24.27 \\
          & g &    07NOV96 & 2.1 m & 4500 & 1.3 & 24.62 \\
    67.0  & r &    06NOV96 & 2.1 m & 1800 & 1.3 & 23.76 \\
          & g &    09NOV96 & 2.1 m & 2700 & 1.5 & 24.10 \\
    79.0  & r &    09NOV96 & 2.1 m & 1260 & 1.7 & 23.26 \\
          & g &    09NOV96 & 2.1 m & 2700 & 1.5 & 24.07 \\
    93.0  & r &    06NOV96 & 2.1 m & 1800 & 1.2 & 23.78 \\
          & g &    08NOV96 & 2.1 m & 3600 & 1.7 & 24.13 \\
    99.0  & r &    05NOV96 & 2.1 m & 1800 & 1.0 & 23.99 \\
          & g &    07NOV96 & 2.1 m & 3600 & 1.3 & 24.61 \\
   109.0  & r &    06NOV96 & 2.1 m & 1800 & 1.6 & 23.54 \\
          & g &    07NOV96 & 2.1 m & 3600 & 1.5 & 24.35 \\
   142.1  & r &    09NOV96 & 2.1 m & 1800 & 1.3 & 23.52 \\
   169.1  & r &    05NOV96 & 2.1 m & 2700 & 0.9 & 24.17 \\
          & g &    08NOV96 & 2.1 m & 4500 & 1.6 & 24.34 \\
   171.0  & r &    07NOV96 & 2.1 m &  600 & 1.0 & 23.24 \\
          & g &    09NOV96 & 2.1 m & 1800 & 1.4 & 24.01 \\
   173.1  & r &    05NOV96 & 2.1 m & 1800 & 1.2 & 23.55 \\
          & g &    09NOV96 & 2.1 m & 2160 & 1.2 & 24.21 \\
   220.1  & r &    05NOV96 & 2.1 m & 2700 & 1.4 & 23.77 \\
          & g &    08NOV96 & 2.1 m & 4500 & 1.8 & 24.12 \\
   303.1  & r &    05JUL94 & 0.9 m & 1800 & 1.3 & 22.61 \\
          & g &    05JUL94 & 0.9 m & 3600 & 1.6 & 23.46 \\
   319.0  & r &    03JUL94 & 0.9 m & 1800 & 1.3 & 22.95 \\
          & g &    03JUL94 & 0.9 m & 3600 & 1.3 & 23.74 \\
   320.0  & r &    02JUL94 & 0.9 m & 2700 & 1.5 & 22.94 \\
          & g &    02JUL94 & 0.9 m & 4500 & 1.5 & 23.60 \\
   323.1  & r &    06JUL94 & 0.9 m & 1800 & 1.3 & 22.68 \\
          & g &    06JUL94 & 0.9 m & 3600 & 1.0 & 23.40 \\
   332.0  & r &    04JUL94 & 0.9 m & 1800 & 1.6 & 22.94 \\
          & g &    04JUL94 & 0.9 m & 3600 & 1.8 & 23.50 \\
   346.0  & r &    04JUL94 & 0.9 m & 1800 & 1.7 & 22.83 \\
          & g &    04JUL94 & 0.9 m & 3600 & 2.0 & 23.40 \\
   348.0  & r &    06JUL94 & 0.9 m & 1800 & 1.8 & 22.79 \\
          & g &    06JUL94 & 0.9 m & 3600 & 1.8 & 23.23 \\
   349.0  & r &    03JUL94 & 0.9 m & 1800 & 1.2 & 23.09 \\
          & g &    03JUL94 & 0.9 m & 3600 & 1.4 & 23.69 \\
   351.0  & r &    09NOV96 & 2.1 m & 1800 & 1.0 & 23.86 \\
          & g &    09NOV96 & 2.1 m & 2700 & 1.1 & 24.26 \\
   357.0  & r &    02JUL94 & 0.9 m & 1800 & 1.4 & 22.98 \\
          & g &    02JUL94 & 0.9 m & 3600 & 1.5 & 23.38 \\
   379.1  & r &    05JUL94 & 0.9 m & 1800 & 1.9 & 22.77 \\
          & g &    05JUL94 & 0.9 m & 3600 & 2.0 & 23.22 \\
   381.0  & r &    03JUL94 & 0.9 m & 1800 & 1.3 & 23.04 \\
          & g &    03JUL94 & 0.9 m & 3600 & 1.5 & 23.38 \\
   401.0  & r &    04JUL94 & 0.9 m & 1800 & 1.6 & 22.91 \\
          & g &    04JUL94 & 0.9 m & 3600 & 1.8 & 23.28 \\
   427.1  & r &    06NOV96 & 2.1 m & 2700 & 1.4 & 23.68 \\
          & g &    07NOV96 & 2.1 m & 3600 & 1.5 & 24.47 \\
   434.0  & r &    05JUL94 & 0.9 m & 2340 & 2.0 & 22.67 \\
          & g &    05JUL94 & 0.9 m & 4320 & 2.2 & 23.29 \\
   436.0  & r &    04JUL94 & 0.9 m & 1800 & 1.8 & 22.63 \\
          & g & 04,06JUL94 & 0.9 m & 2700 & 1.7 & 23.13 \\
   456.0  & r &    08NOV96 & 2.1 m &  900 & 1.1 & 23.25 \\
          & g &    08NOV96 & 2.1 m & 1800 & 1.6 & 23.89 \\
   458.0  & r &    05JUL94 & 0.9 m & 1800 & 1.9 & 22.74 \\
          & g & 05,06JUL94 & 0.9 m & 3600 & 2.0 & 23.01 \\
   459.0  & r &    08NOV96 & 2.1 m &  900 & 1.6 & 23.11 \\
          & g &    08NOV96 & 2.1 m & 1800 & 1.8 & 23.65 \\
   460.0  & r &    03JUL94 & 0.9 m & 2100 & 1.8 & 22.66 \\
          & g &    06JUL94 & 0.9 m & 2700 & 1.9 & 23.07 \\ \tableline
\end{tabular}
\normalsize
\end{table}

\begin{table}
\caption{Radio Galaxy Magnitudes and Colors}
\label{rgmags}
\tiny
\begin{tabular}{cccccccc}
\tableline \tableline
Source & $z$ & Total & Observed & Total &  Total & Nuclear &  Nuclear \\
(3CR\#) & & $m_r$ & $g-r$ &  $M_r$ & $M_g$ & $M_r$ &  $M_B$ \\ \tableline
  16.0  & 0.406  & 20.07  & 1.31 & -22.86  & -22.75  &        &        \\
  17.0  & 0.2197 & 18.05  & 0.54 & -23.07  & -23.06  & -17.67 & -17.53 \\
  18.0  & 0.188  & 18.82  & 0.47 & -21.90  & -21.85  &        &        \\
  19.0  & 0.482  & 19.89  & 1.45 & -23.78  & -23.49  & -18.20 & -18.05 \\
  28.0  & 0.1952 & 17.41  & 0.88 & -23.41  & -22.97  & $>$ -19.09& $>$ -18.95 \\
  42.0  & 0.395  & 19.42  & 1.33 & -23.49  & -23.37  & -19.86 & -19.72 \\
  46.0  & 0.4373 & 18.98  &      & -24.31  &         & -18.11 & -17.97 \\
  49.0  & 0.621  & 20.84  & 1.11 & -24.00  & -23.87  &        &         \\
  61.1  & 0.186  & 20.70: &      &         &         & $>$-18.98&  $>$ -18.83 \\
  63.0  & 0.175  & 18.20: &      & -22.30: &         & -20.42 & -20.28 \\
  67.0  & 0.3102 & 18.86  & 1.07 & -23.31  & -23.18  & -21.08 & -20.94 \\
  79.0  & 0.2559 & 17.62  & 1.20 & -24.08  & -23.63  & -19.45 & -19.30 \\
  93.1  & 0.244  & 19.63: &      & -22.21: &         & -20.21 & -20.07 \\
  99.0  & 0.426  & 18.70  & 0.87 & -24.72  & -25.15  &        &      \\
 109.0  & 0.3056 & 17.10  & 0.93 & -25.53  & -25.72  & -23.82 &-23.68 \\
 142.1  & 0.4061 & 19.97  &      & -23.77  &         &        &       \\
 169.1  & 0.633  & 20.62  & 0.92 & -24.55  & -24.70  &        &      \\
 171.0  & 0.2384 & 18.23  & 0.89 & -23.20  & -22.96  & -20.17 & -20.03 \\
 173.1  & 0.292  & 17.70  & 1.27 & -24.29  & -23.90  & $>$-20.48& $>$-20.34 \\
 196.1  & 0.198  & 18.09: &      & -22.84: &         & -18.45 & -18.30 \\
 200.0  & 0.458  & 20.43: &      & -22.96: &         & $>$-20.41&$>$ -20.27 \\
 213.1  & 0.194  & 16.98: &      & -23.78: &         & -20.18 & -20.03 \\
 219.0  & 0.1744 & 18.34: &      & -22.11: &         & -20.69 & -20.54 \\
 220.1  & 0.620  & 20.55  & 0.89 & -24.28  & -24.37  &        &        \\
 234.0  & 0.1848 & 17.97: &      & -22.67: &         & -17.01 & -16.86 \\
 244.1  & 0.428  & 19.64: &      & -23.41: &         & -17.18 & -17.04 \\
 268.2  & 0.362  & 19.63: &      & -22.85: &         & -19.77 & -19.63 \\
 268.3  & 0.371  & 20.41: &      & -22.15: &         & -19.61 & -19.47 \\
 274.1  & 0.422  & 20.32: &      & -22.74: &         & $>$-18.83&  $>$-18.68 \\
 275.0  & 0.480  & 21.32: &      & -22.22: &         & $>$-19.50& $>$-19.36 \\
 277.0  & 0.414  & 20.24: &      & -22.68: &         & $>$-17.52&  $>$-17.38 \\
 284.0  & 0.2394 & 18.59: &      & -22.72: &         & -20.11 & -19.96 \\
 287.1  & 0.2159 & 18.49: &      & -22.54: &         & -20.72 & -20.58 \\
 288.0  & 0.246  & 17.91: &      & -23.45: &         & -19.29 & -19.15 \\
 299.0  & 0.367  & 21.48: &      & -21.04: &         & -18.96 & -18.81 \\
 300.0  & 0.270  & 18.41: &      & -23.21: &         & -18.82 & -18.67 \\
 303.1  & 0.267  & 18.19  & 1.04 & -23.47  & -23.17  & -20.08 & -19.94 \\
 306.1  & 0.441  & 20.02: &      & -23.30: &         & -20.45 & -20.31 \\
 313.0  & 0.461  & 19.25: &      & -24.13: &         & -23.15 & -23.00 \\
 319.0  & 0.192  & 18.67  & 0.81 & -22.07  & -21.68  & $>$-18.06& $>$-17.92 \\
 320.0  & 0.342  & 18.52  & 1.55 & -23.83  & -23.29  & $>$-18.68&  $>$-18.53 \\
 327.1  & 0.4628 & 20.53: &      & -23.05: &         & -17.99 & -17.85 \\
 332.0  & 0.1515 & 17.03  & 0.63 & -23.10  & -22.78  & -20.90 & -20.75 \\
 341.0  & 0.448  & 21.50: &      & -21.80: &         & -20.34 & -20.20 \\
 346.0  & 0.161  & 17.40  & 0.92 & -23.04  & -22.52  & -20.24 & -20.10 \\
 348.0  & 0.154  & 16.75  & 0.83 & -23.61  & -23.17  & $>$-17.93&  $>$-17.78 \\
 349.0  & 0.205  & 18.28  & 0.97 & -22.59  & -22.07  & -19.91 & -19.76 \\
 357.0  & 0.1664 & 16.73  & 0.93 & -23.71  & -23.16  & -20.32 & -20.17 \\
 379.1  & 0.256  & 17.79  & 1.07 & -23.84  & -23.50  & -20.11 & -19.96 \\
 381.0  & 0.1605 & 17.61  & 0.81 & -22.78  & -22.35  & -20.27 & -20.13 \\
 401.0  & 0.201  & 17.78  & 0.92 & -23.23  & -22.81  & $>$-15.33&  $>$-15.18 \\
 411.0  & 0.467  & 21.81: &      & -22.05: &         & -17.17 & -17.03 \\
 427.1  & 0.572  & 21.28  & 1.81 & -23.52  & -22.87  &        &        \\
 434.0  & 0.322  & 19.26  & 1.40 & -23.10  & -22.72  & $>$-17.17&  $>$-17.03 \\
 435.0A & 0.471  & 22.02: &      & -21.52: &         & -19.73 & -19.58 \\
 436.0  & 0.2145 & 17.91  & 1.02 & -23.35  & -22.93  & -17.69 & -17.54 \\
 456.0  & 0.2330 & 18.20  & 0.88 & -23.09  & -22.80  & -20.84 & -20.70 \\
 458.0  & 0.290  & 19.44  & 1.28 & -22.46  & -22.02  &        &        \\
 459.0  & 0.2199 & 17.43  & 0.21 & -23.71  & -24.04  & -17.92 & -17.77 \\
 460.0  & 0.268  & 18.61  & 0.83 & -23.10  & -23.17  & -19.28 & -19.13 \\ \tableline
\end{tabular}
\normalsize
\end{table}

\begin{table}
\tabletypesize{\scriptsize}
\caption{Quasar Magnitudes and Colors}
\label{qmags}
\tiny
\begin{tabular}{ccccccc}
\tableline \tableline
Source & $z$ & Total & Observed & Total & Total  & Total \\
(3CR\#) & & $m_r$ & $g-r$ &  $M_r$ & $M_g$ & $M_B$ \\ \tableline

  47.0 & 0.425  & 17.90 &  -0.20  & -24.47&  -24.72  & -24.33 \\
  93.0 & 0.358  & 18.29 & 0.49 & -23.95&   -23.60  &  -23.80 \\
 323.1 & 0.264  & 16.21 & 0.14 & -25.01&   -24.90   & -24.86\\
 351.0 & 0.371  & 15.47 & -0.02 & -26.46&  -26.48  & -26.32 \\  \tableline
\end{tabular}
\normalsize
\end{table}

\begin{table}
\tabletypesize{\scriptsize}
\caption{Comparison of Excess Galaxy Measurement Methods}
\label{bggpars}
\tiny
\begin{tabular}{cccccccccc}
\tableline \tableline
Method & Tabulated & $H_0$ & $q_0$ & Optical & Counting & $m_{lim}$ & Background & Stars vs. & Other \\
 & Quantity & (km s$^{-1}$ Mpc$^{-1}$) & & Waveband & Area & & Counts & Galaxies? & Notes \\ \tableline
Ours & $B_{gg}$ & 50 & 0.02 & Gunn $r$ & 0.5 Mpc radius & $m_{comp}$ or $M^*_r(z)$ + 2.5 & YGS+Y & Yes & \\
YE & $B_{gg}$ & 50 & 0.02, 0.5 & Gunn $r$ & 0.5 Mpc radius & $m_{comp}$ or $M^*_r(z)$ + 2.5 & YGS & Yes & 1 \\
Z & $N^{-19}_{0.5}$ & 50 & 0 & V & 0.5 Mpc radius & $M_V$ = -19 & 20$'$-60$'$ offset & Yes & 2 \\
YMP & $B_{gg}$ & 50 & 0 & R$_c$ to Gunn $r$ & variable & $m_{comp}$ & 5$'$-10$'$ offset & No & 3 \\
HL & $N^M_{0.5}$ & 50 & 0.5 & R & 0.5 Mpc radius & $m_1$ + 3 & 30$'$ offset & No & \\ \tableline
\end{tabular}
\normalsize

\tablecomments{
\newline
(1) Some fields done using the luminosity function of Sebok 1986 and $q_0$ = 0.5.
\newline
(2) No galaxies with ``anomalous'' colors.
\newline
(3) Different definition of the completeness limit, $m_{comp}$ from YE and this paper.}

\tablerefs{
\newline
HL: Hill \& Lilly (1991);
\newline
Y: Yee, private communication;
\newline
YE: Yee \& Ellingson (1993);
\newline
YGS: Yee, Green \& Stockman (1986);
\newline
YMP: Yates, Miller \& Peacock (1989);
\newline
Z: Zirbel (1997).}

\end{table}

\begin{table}
\tabletypesize{\scriptsize}
\caption{$B_{gg}$ Values}
\label{allbggs}
\tiny
\begin{tabular}{ccccccccc}
\tableline \tableline
3CR\# & $z$ & Ours & YE & Z & Z2 & HL & YMP & Notes \\ \tableline
16.0 & 0.406 & 439 $\pm$ 151 &  & 502 $\pm$ 232 &  &  & 527 $\pm$ 173 &  \\
17.0 & 0.2197 & 223 $\pm$ 116 &  & 357 $\pm$ 163 &  &  & 93 $\pm$ 90 &  \\
18.0 & 0.188 & -120 $\pm$ 109 &  & -27 $\pm$ 95 &  &  &  &  \\
19.0 & 0.482 & 249 $\pm$ 170 &  &  &  &  &  &  \\
28.0 & 0.1952 & 1179 $\pm$ 110 &  &  &  &  &  &  \\
42.0 & 0.395 & -101 $\pm$ 149 &  & -156 $\pm$ 152 &  &  &  &  \\
46.0 & 0.4373 & 341 $\pm$ 158 &  &  &  &  &  &  \\
47.0 & 0.425 & -119 $\pm$ 157 &  &  &  &  &  &  \\
48.0 & 0.367 & & 108 $\pm$ 118 &  &  &  &  &  \\
49.0 & 0.621 & 232 $\pm$ 252 &  &  &  &  &  &  \\
63.0 & 0.175 &  &  & 441 $\pm$ 163 &  &  &  &  \\
67.0 & 0.3102 & -153 $\pm$ 134 &  & -72 $\pm$ 201 &  &  &  &  \\
79.0 & 0.2559 & 90 $\pm$ 126 &  &  &  &  &  &  \\
93.0 & 0.358 & -252 $\pm$ 144 &  &  &  &  &  &  \\
99.0 & 0.426 & -169 $\pm$ 155 &  &  &  &  &  &  \\
109.0 & 0.3056 & -185 $\pm$ 133 &  & 76 $\pm$ 201 &  &  &  &  \\
142.1 & 0.4061 & 54 $\pm$ 161 &  &  &  &  &  &  \\
169.1 & 0.633 & -312 $\pm$ 302 &  &  &  &  &  &  \\
171.0 & 0.2384 & 57 $\pm$ 118 &  &  &  &  &  &  \\
173.1 & 0.292 & 806 $\pm$ 129 &  &  &  &  &  &  \\
196.1 & 0.198 &  &  & 448 $\pm$ 175 &  &  &  &  \\
200.0 & 0.458 &  &  &  & 38 $\pm$ 228 (Z) & 66 $\pm$ 198 &  &  \\
215.0 & 0.411 & & 1000 $\pm$ 242 &  &  & 1221 $\pm$ 198 &  &  \\
220.1 & 0.620 & 418 $\pm$ 330 &  &  &  &  &  &  \\
225.0B & 0.582 &  &  &  &  &  & 185 $\pm$ 574 & 1 \\
228.0 & 0.5524 &  &  &  &  & 594 $\pm$ 495 & 1230 $\pm$ 468 & 2 \\
244.1 & 0.428 &  &  &  & 950 $\pm$ 376 (Z) & 495 $\pm$ 231 &  & 3 \\
249.1 & 0.311 & & -38 $\pm$ 141 &  &  &  &  &  \\
263.0 & 0.646 & & 993 $\pm$ 550 &  &  &  &  &  \\
268.3 & 0.371 &  &  &  & 304 $\pm$ 228 (Z) & 429 $\pm$ 297 &  &  \\
274.1 & 0.422 &  &  &  & 380 $\pm$ 251 (Z) & 231 $\pm$ 198 & 362 $\pm$ 129 & 4 \\
275.0 & 0.480 &  &  &  & 38 $\pm$ 38 (Z) & 66 $\pm$ 231 & 669 $\pm$ 183 & 5 \\
275.1 & 0.557 & & 1125 $\pm$ 399 &  &  &  &  &  \\
287.1 & 0.2159 &  &  & 76 $\pm$ 99 &  &  &  &  \\
295.0 & 0.4614 &  &  &  & 1102 $\pm$ 376 (Z) & 957 $\pm$ 165 &  & 6 \\
299.0 & 0.367 &  &  & 319 $\pm$ 201 & 152 $\pm$ 228 (AEZO) & 66 $\pm$ 165 &  & 7 \\
303.1 & 0.267 & 115 $\pm$ 128 &  &  &  &  &  &  \\
306.1 & 0.441 &  &  & 334 $\pm$ 201 & 190 $\pm$ 380 (AEZO) & 165 $\pm$ 198 & 529 $\pm$ 115 & 8 \\
313.0 & 0.461 &  &  & 441 $\pm$ 315 & 1064 $\pm$ 532 (AEZO) & 726 $\pm$ 231 &  & 9 \\
319.0 & 0.192 & 494 $\pm$ 109 &  &  &  &  &  &  \\
320.0 & 0.342 & 856 $\pm$ 145 &  & 654 $\pm$ 243 &  &  &  &  \\
323.1 & 0.264 & 268 $\pm$ 127 & 440 $\pm$ 224 &  &  &  &  &  \\
327.1 & 0.4628 &  &  &  &  &  & -195 $\pm$ 181 & 10 \\
330.0 & 0.550 &  &  &  &  & 660 $\pm$ 231 &  &  \\
332.0 & 0.1515 & 306 $\pm$ 101 &  & 236 $\pm$ 160 &  &  &  &  \\
334.0 & 0.555 & & 187 $\pm$ 220 &  &  &  &  &  \\
341.0 & 0.448 &  &  & 376 $\pm$ 236 & 266 $\pm$ 380 (AEZO) & 330 $\pm$ 231 &  &  \\
345.0 & 0.594 & & 773 $\pm$ 297 &  &  &  &  &  \\
346.0 & 0.161 & 641 $\pm$ 104 &  & 380 $\pm$ 175 &  &  & 405 $\pm$ 234 &  \\
348.0 & 0.154 & 614 $\pm$ 101 &  & 1201 $\pm$ 281 &  &  & 174 $\pm$ 292 & 11 \\
349.0 & 0.205 & 206 $\pm$ 112 &  &  &  &  &  &  \\
351.0 & 0.371 & -82 $\pm$ 146 & -198 $\pm$ 138 &  &  &  &  &  \\
357.0 & 0.1664 & 370 $\pm$ 103 &  &  &  &  &  &  \\
379.1 & 0.256 & -136 $\pm$ 126 &  &  &  &  &  &  \\
381.0 & 0.1605 & 138 $\pm$ 103 &  & 323 $\pm$ 201 &  &  &  &  \\
401.0 & 0.201 & 1131 $\pm$ 112 &  &  &  &  &  &  \\
427.1 & 0.572 & -166 $\pm$ 289 &  &  &  &  &  &  \\
434.0 & 0.322 & 135 $\pm$ 146 &  & 395 $\pm$ 277 &  &  & 143 $\pm$ 152 &  \\
435.0A & 0.471 &  &  & 1273 $\pm$ 308 &  &  & 573 $\pm$ 227 & 12 \\
436.0 & 0.2145 & 420 $\pm$ 115 &  &  &  &  &  &  \\
455.0 & 0.5427 & & 764 $\pm$ 366 &  &  &  &  &  \\
456.0 & 0.2330 & -85 $\pm$ 116 &  &  &  &  & 107 $\pm$ 183 &  \\
458.0 & 0.290 & 101 $\pm$ 129 &  &  &  &  & -56 $\pm$ 143 &  \\
459.0 & 0.2199 & 33 $\pm$ 116 &  & 99 $\pm$ 95 &  &  &  &  \\
460.0 & 0.268 & -110 $\pm$ 128 &  &  &  &  &  &  \\ \tableline
\end{tabular}
\normalsize

\tablecomments{
\newline
(1) The YMP counting area (A$_{YMP}$) is 1.4 $\times$ the area given by a 0.5 Mpc radius (A$_{r=0.5 Mpc}$).
\newline
(2) The error bars overlap but the difference between the two values is substantial.
The HL value is uncertain due to a large ($>$ a factor of 2) correction due to a ``bright'' m$_{lim}$.
The YMP m$_{lim}$ is even brighter and A$_{YMP}$ = 1.3 $\times$ A$_{r=0.5 Mpc}$.
\newline
(3) The error bars overlap but the difference between the two values is substantial.
HL suggest the comparison field had higher counts than it should have.
\newline
(4) The error bars overlap.  HL m$_{lim}$ = 22.  YMP m$_{lim}$ = 23.51.
A$_{YMP}$ = 2.6 $\times$ A$_{r=0.5 Mpc}$.
\newline
(5) The two smaller values agree with each other but are highly discrepant with the larger
value.  HL m$_{lim}$ = 23.  YMP m$_{lim}$ = 22.88.  A$_{YMP}$ = 2.3 $\times$ A$_{r=0.5 Mpc}$.
\newline
(6) The error bars overlap.  HL mention this cluster is incompletely counted at the M$_B$ = -19.0 level.
\newline
(7) The error bars overlap.  HL m$_{lim}$ = 23.  Z m$_{lim}$ = 23.1.  Z image is in V.
\newline
(8) YMP is discrepant with HL.  HL m$_{lim}$ = 23.  YMP m$_{lim}$ = 23.68.
A$_{YMP}$ = 2.6 $\times$ A$_{r=0.5 Mpc}$.  Z m$_{lim}$ = 23.55.  Z image is in V.
\newline
(9) The error bars overlap but the difference between the values is substantial.  Z states
that a large correction ($>$ 20\%) was needed due to incompleteness at the M$_V$ = -19.0 level.  HL m$_{lim}$ = 23.
Z m$_{lim}$ = 23.7.  Z image is in V.
\newline
(10) YMP state that no offset frame was available for this field and so an offset frame
from another field was used.  A$_{YMP}$ = 1.6 $\times$ A$_{r=0.5 Mpc}$.
\newline
(11) All three values are highly discrepant.  Ours is $\sim$ the average of the other two.
A$_{YMP}$ = 0.34 $\times$ A$_{r=0.5 Mpc}$.  Our m$_{lim}$ = 22.8.  Z m$_{lim}$ = 21.  Z image is in V.
\newline
(12) The two values are highly discrepant.  YMP m$_{lim}$ = 23.13.  A$_{YMP}$ = 1.6
$\times$ A$_{r=0.5 Mpc}$.  Z m$_{lim}$ = 23.7.  Z image is in V.}

\tablerefs{
\newline
AEZO: Allington-Smith et al.\ (1993);
\newline
HL: Hill \& Lilly (1991);
\newline
YE: Yee \& Ellingson (1993);
\newline
YMP: Yates, Miller \& Peacock (1989);
\newline
Z: Zirbel (1997);
\newline
Z2(AEZO): Values of Hill \& Lilly (1991) converted into Zirbel values by Allington-Smith et al.\ (1993);
\newline
Z2(Z): Values of Hill \& Lilly (1991) converted into Zirbel values by Zirbel (1997).}

\end{table}

\begin{table}
\caption{Adopted $B_{gg}$ Values}
\label{finbggs}
\tiny
\begin{tabular}{ccccccc}
\tableline \tableline
Source & $z$ & ID & B$_{gg}$ & $\sigma$ & References & Notes \\
(3CR\#) & & & (Mpc$^{1.77}$) & (Mpc$^{1.77}$) & & \\ \tableline
  16.0 & 0.406  & GAL &  439 & 151 &  & \\
  17.0 & 0.2197 & N   &  223 & 116 &  & \\
  18.0 & 0.188  & GAL & -120 & 109 &  & \\
  19.0 & 0.482  & GAL &  249 & 170 &  & \\
  28.0 & 0.1952 & GAL & 1179 & 110 &  & \\
  42.0 & 0.395  & GAL & -101 & 149 &  & \\
  46.0 & 0.4373 & GAL &  341 & 158 &  & 1 \\
  47.0 & 0.425  & QSO & -119 & 157 &  & \\
  48.0 & 0.367  & QSO &  108 & 118 & YE & \\
  49.0 & 0.621  & GAL &  232 & 252 &  & \\
  63.0 & 0.175  & GAL &  441 & 163 & Z & \\
  67.0 & 0.3102 & GAL & -153 & 134 &  & \\
  79.0 & 0.2559 & N   &   90 & 126 &  & \\
  93.0 & 0.358  & QSO & -252 & 144 &  & \\
  99.0 & 0.426  & N   & -169 & 155 &  & \\
 109.0 & 0.3056 & N   & -185 & 133 &  & \\
 142.1 & 0.4061 & GAL &   54 & 161 &  & 2 \\
 169.1 & 0.633  & GAL & -312 & 302 &  & \\
 171.0 & 0.2384 & N   &   57 & 118 &  & \\
 173.1 & 0.292  & GAL &  806 & 129 &  & 3 \\
 196.1 & 0.198  & GAL &  448 & 175 & Z & \\
 200.0 & 0.458  & GAL &   52 & 213 & avg. of Z2(Z) and HL & \\
 215.0 & 0.411  & QSO & 1000 & 242 & YE & \\
 220.1 & 0.620  & GAL &  418 & 330 &  & \\
225.0B & 0.582  & GAL &  185 & 574 & YMP & \\
 228.0 & 0.5524 & GAL &  912 & 482 & avg. of HL and YMP & 4 \\
 244.1 & 0.428  & GAL &  723 & 304 & avg. of Z2(Z) and HL & 4 \\
 249.1 & 0.311  & QSO &  -38 & 141 & YE & \\
 263.0 & 0.646  & QSO &  993 & 550 & YE & \\
 268.3 & 0.371  & GAL &  367 & 263 & avg. of Z2(Z) and HL & \\
 274.1 & 0.422  & GAL &  324 & 193 & avg. of Z2(Z), HL and YMP & \\
 275.0 & 0.480  & GAL & 38-669 &     & Z2(Z), YMP & 5 \\
 275.1 & 0.557  & QSO & 1125 & 399 & YE & \\
 287.1 & 0.2159 & N   &   76 &  99 & Z & \\
 295.0 & 0.4614 & GAL & 1030 & 271 & avg. of Z2(Z) and HL & \\
 299.0 & 0.367  & GAL &  179 & 198 & avg. of Z, Z2(AEZO) and HL & \\
 303.1 & 0.267  & GAL &  115 & 128 &  & \\
 306.1 & 0.441  & GAL &  305 & 224 & avg. of Z, Z2(AEZO), HL and YMP & 6 \\
 313.0 & 0.461  & GAL &  744 & 359 & avg. of Z, Z2(AEZO) and HL & 4 \\
 319.0 & 0.192  & GAL &  494 & 109 &  & \\
 320.0 & 0.342  & GAL &  856 & 145 &  & \\
 323.1 & 0.264  & QSO &  268 & 127 &  & \\
 327.1 & 0.4628 & GAL & -195 & 181 & YMP & \\
 330.0 & 0.550  & GAL &  660 & 231 & HL & \\
 332.0 & 0.1515 & GAL &  306 & 101 &  & \\
 334.0 & 0.555  & QSO &  187 & 220 & YE & \\
 341.0 & 0.448  & GAL &  324 & 282 & avg. of Z, Z2(AEZO) and HL & \\
 345.0 & 0.594  & QSO &  773 & 297 & YE & \\
 346.0 & 0.161  & GAL &  641 & 104 &  & \\
 348.0 & 0.154  & GAL &  614 & 101 &  & 7 \\
 349.0 & 0.205  & GAL &  206 & 112 &  & \\
 351.0 & 0.371  & QSO &  -82 & 146 &  & \\
 357.0 & 0.1664 & GAL &  370 & 103 &  & \\
 379.1 & 0.256  & GAL & -136 & 126 &  & \\
 381.0 & 0.1605 & GAL &  138 & 103 &  & \\
 401.0 & 0.201  & GAL & 1131 & 112 &  & \\
 427.1 & 0.572  & GAL & -166 & 289 &  & \\
 434.0 & 0.322  & GAL &  135 & 146 &  & \\
435.0A & 0.471  & GAL & 573-1273 &     & YMP, Z & 8 \\
 436.0 & 0.2145 & GAL &  420 & 115 &  & \\
 455.0 & 0.5427 & QSO &  764 & 366 & YE & \\
 456.0 & 0.2330 & GAL &  -85 & 116 &  & \\
 458.0 & 0.290  & GAL &  101 & 129 &  & \\
 459.0 & 0.2199 & N   &   33 & 116 &  & \\
 460.0 & 0.268  & GAL & -110 & 128 &  & \\ 
\end{tabular}
\normalsize

\tablecomments{
\newline
(1) We have no g frames for this field.  There is a very bright star near the object.
\newline
(2) We have no g frames for this field.  The ID of this object is uncertain.
\newline
(3) There is a very bright star near the object.
\newline
(4) Although the error bars for the different $B_{gg}$ values overlap, the $B_{gg}$
values themselves differ by more than an Abell richness class (400 $B_{gg}$ units).
\newline
(5) The Z2(Z) and HL values agree with each other but are inconsistent with the
value of YMP which is greater than these values by more than an Abell richness class.
\newline
(6) The values of HL and YMP are not consistent with one another.  However, these
values differ by less than an Abell richness class and both are in agreement with
the values of Z and Z2(AEZO).  Thus, all values are used in the calculation of the
average.
\newline
(7) Values from the literature do not agree with our value.  The value of Z is
greater than our value by more than an Abell richness class while the value of YMP
is less than our value by more than an Abell richness class.  Our value is nearly
the average of these other two values.
\newline
(8) The values of Z and YMP are not consistent with one another and differ by more
than an Abell richness class.}

\tablerefs{
\newline
AEZO: Allington-Smith et al.\ (1993);
\newline
HL: Hill \& Lilly (1991);
\newline
YE: Yee \& Ellingson (1993);
\newline
YMP: Yates, Miller \& Peacock (1989);
\newline
Z: Zirbel (1997);
\newline
Z2(AEZO): Values of Hill \& Lilly 1991 converted into Zirbel values by Allington-Smith et al.\ (1993);
\newline
Z2(Z): Values of Hill \& Lilly 1991 converted into Zirbel values by Zirbel (1997).}

\end{table}

\end{document}